\documentclass[%
 reprint,
nofootinbib,
 amsmath,amssymb,
 aps,
]{revtex4-1}

\usepackage[caption=false]{subfig}

\usepackage{graphicx}
\usepackage{dcolumn}
\usepackage{bm}
\usepackage{amsmath}
\usepackage{mathtools}

\newcommand\numberthis{\addtocounter{equation}{1}\tag{\theequation}}

\begin{document}


\title{Geometry of criticality, supercriticality and Hawking-Page transitions in Gauss-Bonnet-AdS black holes}

\author{Anurag Sahay}
 \altaffiliation{sahay.anurag12@gmail.com}
 \affiliation{%
 Department of Physics, National Institute of Technology Patna, Patna 800005, India
}%
\author{Rishabh Jha}%
\altaffiliation{rishabh.jha77@gmail.com}
 \affiliation{%
 Simulate Learning Solutions Private Limited, Kanpur, Uttar Pradesh, 208017, India
}%

\begin{abstract}
We obtain the Ruppeiner geometry associated with the non-extended state space ($\Lambda$ constant) of the charged Gauss-Bonnet AdS (GB-AdS) black holes and confirm that the state space Riemannian manifold becomes strongly curved in regions where the black hole system develops strong statistical correlations in the grand canonical ensemble ($M$ and $Q$ fluctuating). We establish the exact proportionality between the state space scalar curvature $R$ and the inverse of the singular free energy near the isolated critical point for the grand canonical ensemble in spacetime dimension $d=5$, thus hopefully moving a step closer to the agenda of a concrete physical interpretation of $R$ for black holes. On the other hand, we show that while $R$ signals the Davies transition points (which exist in GB-AdS black holes for $d \ge 6$) through its divergence, it does not scale as the inverse of the singular free energy there. Furthermore, adapting to the black hole case the ideas developed in \cite{rupp2} in the context of pure fluids, we find that the state space geometry encodes phase coexistence and first order transitions, identifies the asymptotically critical region and even suggests a Widom line like crossover regime in the supercritical region for $5-d$ case. The sign of $R$ appears to imply a significant difference between the microscopic structure of the small and the large black hole branches in $d=5$. We show that thermodynamic geometry informs the microscopic nature of coexisting thermal GB-AdS and black hole phases near the Hawking-Page phase transition. 
\end{abstract}

\pacs{Valid PACS appear here}
\maketitle


\section{Introduction}
Einstein gravity is naturally understood as a low energy limit of some UV complete quantum theory of gravity. Higher order curvature corrections to the Einstein-Hilbert action are expected to perturbatively increase the gravitational strength  towards the quantum regime. Symbolically, the gravity Lagrangian appears as an expansion in the  coupling constant $\alpha$ of the higher derivative terms,
\begin{align*}
\mathcal{L}=R+\alpha' R^2+\alpha'^2 R^3+\ldots
\end{align*}
where the powers of $R$ in each term are representative of all combinations of the Riemann tensor. A generic addition of higher derivative terms however leads to undesirable features like negative energy modes or ghosts. Fortunately, it turns out, specific combinations at each order of the higher order curvature terms produce ghost free and tractable solutions which are only second order in the metric. In this unique combination of higher derivative terms (with arbitrary coefficients) known as the Lovelock Lagrangian, each of the $R^n$ term represents the Euler density for spacetime of dimension $d=2 n$ \citep{love}. 
 
A truncation of the Lovelock theory to the leading  order quadratic term is known as the Gauss-Bonnet (GB) gravity and it lends itself to explicit gravity solutions. In fact, it is the only consistent quadratic correction to Einstein gravity \citep{deser}. Since the Gauss-Bonnet correction term is an Euler density and hence a topological invariant in $d=4$, non trivial dynamics appears only for $d\geq 5$. A study of the GB terms finds further motivation from string theory. The GB corrections automatically emerge as the ghost-free one loop terms of order $\alpha'$ (the string length scale squared) in the low energy effective actions of some string theories \cite{deser,zwei}, with the tree level term in the expansion beingt the Einstin-Hilbet action. Significantly, string theory supplies yet another perspective to the GB corrections to Einstein graviy in anti-de Sitter (AdS) spacetime via the celebrated AdS/CFT correspondence.

 The AdS/CFT correspondence first explicitly stated as a conjecture by Maldacena \citep{malda}, is a strong/weak duality between a theory with gravity in AdS spacetime and a gravity-free gauge theory in one dimension less, sometimes said to be living on the boundary of AdS. One end of the weak/strong spectrum which has been extensively studied relates a weakly coupled gravity theory, namely Einstein gravity, to a strongly coupled gauge theory. This has proven to be a computation bonanza as, spurred on by the conjecture, researchers have come up with ingenious calculations to confirm its validity and make some general, universal predictions about the mostly intractable strongly coupled field theory end of the duality. Thus, for example, it was shown in the early days of the conjecture that the well known Hawking-Page transition in the AdS black holes is in fact dual to a confinement-deconfinement transition in the strongly coupled dual gauge theory \citep{witt}. The duality conjecture has stimulated intense activity over the past several years in various sub-fields like AdS/QCD \cite{kovtun}, Fluid/Gravity correspondence  \cite{mukund}, AdS/CMT \cite{subir}, holographic entanglement entropy (HEE)  \cite{ryu}, etc.
 
  In the context of the duality, loading the AdS gravity with the Gauss-Bonnet coupling would correspond to reducing the coupling strength of the dual gauge theory, namely placing it in the regime of finite $N$, the order of the gauge group, and finite $\lambda$, the 't Hooft coupling\cite{odin,sudipt1,sudipt2}. It turns out, the technical  issues in this latter regime of duality are not as well ironed out as the former case and many possibilities remain to be explored \citep{thesis}. A thorough investigation of the thermodynamics and phase behaviour of the Gauss-Bonnet AdS black holes would therefore go a long way in shedding light on this regime. Indeed, thermodynamics in a GB setup has been extensively studied over the past years. It is well known that the Gauss-Bonnet corrected AdS black holes still undergo Hawking-Page transition to the thermal Gauss-Bonnet AdS, if only for the spherical horizon topology. These black holes display a rich phase behaviour which is often qualitatively different from their tree level counterparts. The phase structure and thermodynamics of Gauss-Bonnet AdS black holes have been investigated in different ensembles, in both the non-extended and the extended state space, and with additional fields like the dilaton, Maxwell or even higher order correction to the gauge fields \cite{odin}-\cite{greek}.  A phenomenological dual matrix model has also been proposed which qualitatively reproduces the phases of both the uncharged and the $R$-charged Gauss-Bonnet AdS black holes  \cite{sudipt1,sudipt2}. Recently, a quasinormal mode analysis of GB-AdS black holes from a holographic viewpoint has been undertaken \cite{konop}.

 In recent times there has been an active interest in the information geometry associated with the thermodynamics of black holes \cite{jan, ruppm1}. 
  It is an extension to the black hole thermodynamic system of the thermodynamic geometry (TG) of ordinary extensive systems. TG in its most widely used form was pioneered by Ruppeiner \cite{rupp0,rupp} and its extension to black hole systems was first considered by Gibbons, Kallosh and Kol, \cite{kall}. Ruppeiner's thermodynamic geometry is a covariant theory of thermodynamic fluctuations wherein the thermodynamic state space is uplifted to a Riemannian manifold by incorporating the negative Hessian of the entropy as the metric. The probability of a spontaneous fluctuation between two nearby equilibrium states decreases as the exponential of the negative invariant distance measure between the two points in the thermodynamic manifold. The thermodynamic manifold responds to the underlying statistical correlations in the system. It is curved for an interacting system and flat for a non-interacting one. Moreover, in regions of very strong correlations, like the critical point, the geometry becomes singular. Furthermore, the nature of the curved thermodynamic  manifold, whether hyperbolic or elliptic, is apparently decided by the nature of the underlying statistical correlations, whether long range attractive or short range repulsive.  For extensive systems the invariant scalar curvature $R$ turns out to be proportional to the correlation volume $\xi^d$ where $d$ is the dimensionality of the system. At a more fundamental level, the scalar curvature $R$ is conjectured to be proportional to the inverse of the singular part of the free entropy(or Massieu function)  $\psi_s$ near the critical point. Following \cite{rupp} it is conventional to show the proportionality of $R$ with the inverse of the singular part of free energy density,
\begin{align}
  R=-\kappa \frac{1}{\Phi_s}
  \label{Rvsphi1}
\end{align}
  where $\kappa$ is positive and order unity. Here,  $\Phi_s$ is the singular part of the free energy density near the critical point. This explains the connection of $R$ with the correlation volume since near criticality $\psi_s^{-1}$ is proportional to $\xi^d$.  Equation (\ref{Rvsphi1}) above is reminiscent of Einstein's equation which relates the spacetime curvature to the matter-energy distribution. 
  
  Significantly, in the past few years various investigations in the context of fluids have established that thermodynamic geometry effectively encodes the phase coexistence region as well as the supercritical regime of pure fluids in the vicinity of the critical point. The correspondence of $R$ with the correlation length makes it eminently possible to implement in a thermodynamic set up Widom's arguments about the equality of the length scale of density fluctuations, and hence the correlation length, in the coexisting phases. It has been generally verified with NIST data for fluids and confirmed through extensive simulations of Lennard Jones fluids that $R$ of the coexisting phases is equal, thus providing an alternative to the Maxwell's rule for obtaining phase coexistence \citep{rupp2, rupp3}.  In addition, $R$ uniquely identifies the Widom line crossover region defined as the region of maximum correlation length in the supercritical region.
  
  The agenda of the program of extension of TG to black holes is to bring the useful features of the TG of ordinary systems to bear upon the black hole thermodynamic system. This is especially imperative since the knowledge of black hole microscopics is a work in progress despite some stellar achievements by string theory in calculating black hole microstates. A partial list of the extensive literature on the TG of black holes can be seen in the reference section of \cite{sahay}.
  
  In this paper we address the information geometry of the Einstein-Maxwell Gauss-Bonnet AdS black hole thermodynamics in $d=5$ and $d=6$. As one of us had emphasized in \citep{sahay}, the thermodynamic geometry of the system in which the Hessian of the entropy has been taken with respect to all the available extensive variables by construction represents the system in the grand canonical ensemble. Since the black hole systems we describe in this paper have only two thermodynamic degrees of freedom, the ADM mass $M$ and the charge $Q$, the only possible TG for the system represents the grand canonical ensemble in which the both the mass and the charge are allowed to fluctuate. Note that in this paper we restrict our investigation to the non-extended thermodynamic state space wherein the cosmological constant is a fixed parameter. Gauss-Bonnet black holes have been investigated in the extended state space and $P-V$ criticality has been reported for both the canonical as well as the grand canonical ensembles \cite{caicaicai,zou}. In fact, for the latter case there is a line of critical points. However, since for the static black holes, charged or uncharged, the thermodynamic volume has a monotonic relation with the entropy, it is not possible to obtain a TG with independent volume fluctuations \citep{dolanT}. Therefore, early on in the paper we shall scale away the cosmological constant. A unique feature of the Gauss-Bonnet corrections is that in the grand canonical ensemble the $d=5$ black holes have phase coexistence regions leading to an isolated critical point. Analogous to the $P-V$ criticality in the extended space the non-extended state space features a $\Phi-Q$ criticality where $Q$ is the order parameter while $\Phi$ is the `ordering field'. The critical exponents again turn out to be mean field as before.
  
   Significantly for TG, phase coexistence and critical phenomena feature in the grand canonical ensemble of the $d=5$ GB-AdS black hole thus opening up the system to the full range of the TG treatment. We try not to miss out on this opportunity in this work. Indeed, we find that the thermodynamic manifold is singular at the critical  point as well as the spinodal points. Moving beyond this routine description of $R$, we find that quite to our satisfaction, we are able to deliver good on our promise made earlier in \citep{sahay} that just as for ordinary systems the non-dimensional scalar curvature $R$ for the black holes is proportional to the singular part of the free energy near the critical point. We show this explicitly for the $d=5$ case, namely, 
  
\begin{equation}
 R_{bh}=-\kappa\frac{1}{\Phi_s}
\end{equation}

 where, $\Phi_s$ is the free energy of the black hole and $\kappa$ however is no more an order unity constant but a function only of the coupling constant $\alpha$. And yet, we find that $R$ does not scale as the singular free energy for the Davies transitions in the $d=6$ case even as it does diverge there. We then numerically study the geometry in the phase coexistence region as well as the supercritical region for $d=5$ and again find that, analogous to the pure fluid case, the scalar curvature nicely describes the vicinity of the critical point. 
 
 We note here that a previous work on the thermodynamic geometry of the charged GB-AdS black holes has appeared in \cite{gbc1}. There the authors use the Hessian of a different thermodynamic potential as the starting point even though they continue to address it as the Ruppeiner metric. Their geometry has a singularity at the critical point of the canonical ensemble. In a subsequent section we shall present our arguments disfavoring a geometric investigation of the canonical ensemble.
 
 This paper is organized as follows. In section II we introduce the charged Gauss-Bonnet AdS black holes and their thermodynamic quantities. In subsection IIA, we give a brief overview of phase structure, in subsection IIB we describe the phase structure for the $d=5$ case and in subsection IIC the $d=6$ case. For each case we also discuss the scaling behaviour near the singular points. In section III, we discuss the thermodynamic geometry first for the case of $d=5$ and then for the case of $d=6$. Finally we present our conclusions.

\section{Gauss-Bonnet AdS Black Holes and their Thermodynamics}
The $d$-dimensional Einstein-Maxwell Gauss-Bonnet action in AdS space is given as follows \citep{sudipt1,sudipt2,cai}.
\begin{align*}
S &= S_g+S_m\\ &=\int d^dx\,\sqrt{-g}\,\biggl[\frac{1}{16\pi\,G_d}\bigl(R-2\,\Lambda + \alpha(R^2-  \\  &4\,R_{\mu\nu}\,R^{\mu\nu}+\,R_{\mu\nu\rho\sigma}\,R^{\mu\nu\rho\sigma})\bigr)-\frac{1}{4}\,F_{\mu\nu}\,F^{\mu\nu}\biggr]\numberthis \label{action}
\end{align*}

It is usual to assume the Gauss-Bonnet(GB) coupling constant  $\alpha $ to be positive since GB corrections to gravity are well motivated from string theory where they naturally occur in the $\alpha '$ expansions of the low energy effective action, with $\alpha$ proportional to $\alpha '$, the string length squared \cite{deser}. Moreover, since the GB term is the Euler density and hence a topological invariant in $d=4$, we shall work with spacetimes with $d\geq 5$. The cosmological constant $\Lambda$ is expressed in terms of the AdS length as $\Lambda = -(d-1)(d-2)/2\,l^2$. The matter content of the action comes from the gauge field strength tensor $F_{\mu\nu}=\partial_{\mu} A_{\nu}-\partial_{\nu} A_{\mu}$. The $d$-dimensional gravitational constant $G_d$ shall be set to unity henceforth. 

The field equations are obtained by varying the action and the  general static solution of a charged Gauss-Bonnet-AdS (GB-AdS) is given by the following metric \cite{cai,greek}.
\begin{equation}
ds^{2}=-f(r)dt^{2}+\frac{1}{f(r)}dr^{2}+r^{2}h_{ij}dx^{i}dx^{j}
\label{metric}
\end{equation}\\
and the gauge field
\begin{equation}
A_{t}=-\frac{Q \Sigma  }{16 \pi  (d-3)r^{d-3}}+ \Phi
\label{gauge}
\end{equation}
where $\Phi$ is a constant. Here, $h_{ij}$ is the metric of the maximally symmetric Einstein space with the constant curvature $(d-2)(d-3)k$. Thus the curvature parameter $k$ can be taken to be simply $k=+1,0$ and $-1$ giving a spherical, flat and a hyperbolic horizon topology respectively. The metric function $f$ is given as,

\begin{widetext}
\begin{equation}
f=k+\frac{r^2}{2 (d-4) (d-3) \alpha } \left(1-\sqrt{1-\frac{4 (d-4) (d-3) \alpha }{l^2}-\frac{2 (d-4) Q^2 \alpha }{(d-2)r^{2 d-4} }+\frac{64 (d-4) (d-3) M \pi \text{  }\alpha }{(d-2) \Sigma_{d-2}  r^{d-1}}}\right)
\end{equation}
\end{widetext}

Here, $\Sigma_{d-2}$ is the surface area of the $(d-2)$ dimensional Einstein surface of unit radius, $M$ is the ADM mass and $Q$ the charge of the black hole. The $M=Q=0$ solution, namely the pure Gauss-Bonnet AdS spacetime imposes the following constraint on $\alpha$,
\begin{align}
0 \leq \,\, \frac{\alpha}{l^2}\, \leq\, \frac{1}{4(d-3)(d-4)}
\end{align} 	 

In this paper we shall work with the spherical black hole solutions, namely the $k=+1$ case. This is simply because they are the most interesting from the point of view of phase coexistence and criticality which is the main focus of this work. The details of thermodynamics and phase structures for the $k=0$ and $k=-1$ cases can be found in \cite{cai, caicai, greek}. 

The mass of the spherical black hole can be obtained in terms of the charge and the horizon radius from the horizon condition $f(r)=0$.
\begin{widetext}
\begin{equation}
\label{mass}
M = \frac{\left(2 \left(d^2-5 d+6\right) r^{2(d+2)}+l^2 \left(Q^2 r^8+2 \left(d^2-5 d+6\right) r^{2 d} \left(r^2+\left(d^2-7 d+12\right) \alpha \right)\right)\right) \Sigma }{32 (-3+d) l^2 \pi r^{d+5}}\\
\end{equation}
\end{widetext}

Note that in the following, we shall abuse the symbol $`r'$ to make it stand for the horizon radius $r_h$ which solves the horizon equation $f(r_h)=0$

The quantity $\Phi$ at the horizon is the potential difference between the horizon and infinity eq. (\ref{gauge}),
\begin{equation}
\label{phi}
\Phi = \frac{Q \Sigma  }{16 \pi  (d-3)r^{d-3}}
\end{equation}

Temperature can be straightforwardly obtained by demanding the absence of conical defect singularity at the horizon in the Euclidean metric, thereby giving the following relation.
\begin{widetext}
\begin{equation}
\label{temp}
T = \frac{\left(2 \left(d^2-3d+2\right) r^{4+2 d}+l^2 \left(-Q^2 r^8+2 \left(d^2-5d+6\right) r^{2 d} \left(r^2+\left(d^2-9d+20\right) \alpha \right)\right)\right)}{8 (d-2) \left(r^2+2 \left(d^2-7d+12\right) \alpha \right) l^2 \pi r^{2d+1} }
\end{equation}
\end{widetext}

The entropy is given by the Bekenstein-Hawking one-fourth area term plus a correction term arising on account of the Gauss-Bonnet coupling constant. It is given as follows:
\begin{equation}
\label{entropy}
S =\int_0^r \frac{1}{T}\frac{\partial M}{\partial r}\,dr=\frac{1}{4} \Sigma  r^{d-2} \left(1+\frac{2 \alpha  (d-2)}{(d-4) r^2}\right)
\end{equation}

The Gibbs free energy of the black hole, $G=M-TS-\Phi Q$
can similarly be obtained from the above equations. One can obtain the Gibbs free energy in a more fundamental manner by calculating the Euclidean action of gravity with a fixed gauge potential at the boundary and subtracting from it the Euclidean action of the thermal Gauss-Bonnet AdS with the same value of the pure gauge potential at the boundary \citep{sudipt1,sudipt2,cho}. The normalized Euclidean action $\mathcal{I}$ thus obtained directly relates to the Gibbs free energy,
\begin{equation}
G(\beta,\Phi) =\beta\, \mathcal{I}(\beta,\Phi)
\label{eucidean}
\end{equation}
Starting with the Gibbs free energy one can in turn obtain the  thermodynamic variables $M,Q$ and $S$. At negative values of $G$ the black hole phase dominates thermal AdS while at positive values of $G$ it is the other way round. $G=0$ therefore marks the Hawking-Page phase transition.

Of late there has been an active interest in treating the cosmological constant $\Lambda$ as a thermodynamic variable. This fits well with the understanding that the cosmological constant is proportional to the pressure, $P=-8\pi/\Lambda$. The thermodynamic volume conjugate to the pressure term is obtained via the relation $V = \frac{\partial M}{\partial P}|_{S, Q}$, 
\begin{equation}
V = \frac{\Sigma\,  r^{d-1}}{d-1}
\end{equation}

As mentioned previously, it can be seen that, just like the entropy, the thermodynamic volume depends only upon the horizon radius, $r_h$, and hence the two quantities are not independent of each other.
Just as for any other static metric like the Reissner-Nordstr$\ddot{o}$m-AdS (RN-AdS) or the Schwarzschild-AdS case the thermodynamic volume of the GB-AdS black hole too is not independent of its entropy. Thus, for example its heat capacity at constant volume is zero. While this is useful for, say, implementing holographic heat engines \cite{holoheatengine, chandra}, it is not possible to have thermodynamic ensembles of black holes in which volume fluctuations are independent of other extensive quantities. Unfortunately for TG, therefore, one cannot  incorporate volume (and hence pressure) fluctuations in the state space manifold of GB-AdS black holes and so TG only addresses the non-extended thermodynamics for static black holes, including the GB-AdS case at hand.

With the constraint of constant $P$ for the non-extended phase space the first law is written more transparently in terms of the enthalpy rather than the internal energy,
\begin{align*}
dS&=\frac{1}{T}\,dE+\frac{P}{T}\,dV-\frac{\Phi}{T}\,dQ\\
&=\frac{1}{T}\,d(E-PV)-\frac{\Phi}{T}\,dQ\\
&=\frac{1}{T}\,dM-\frac{\Phi}{T}\,dQ\numberthis
\end{align*},
where $E=M-PV$ is the internal energy. Now, $M$ and $Q$ are independent fluctuations.
 Since we are not going to vary the pressure, in the following we shall set the AdS length scale to unity and continue to use the same notations for all the scaled variables like the GB coupling, mass, temperature, etc.

\subsection{Phase Structure}

It is well known that spherical Gauss-Bonnet AdS black holes in $d=5$ display phase coexistence and criticality between the locally stable small black hole and the large black hole branches in both the grand canonical and the canonical ensembles for the charged case and for the canonical ensemble of the uncharged case as well  \cite{cai, sudipt1,sudipt2,cho}. For $d\geq 6$ onward the phase coexistence phenomena disappears in the grand canonical ensemble of the charged black hole as well as for the uncharged case, and is replaced by the Davies `transition' between the stable large black hole and the unstable small black hole branches. Phase coexistence and criticality continue to manifest in the canonical ensemble of the charged $d\geq 6$ black holes. We must point out that it is very rare indeed for the black holes in the grand canonical ensemble to exhibit phase coexistence and criticality, which is the case for $d=5$ charged GB-AdS, the focus of our present work. This is especially significant for the thermodynamic geometry investigations of black holes because at least two dimensions are necessary for a non-trivial thermodynamic manifold. In other words, one needs two fluctuating conserved variables of the thermodynamic system, which immediately chooses grand canonical ensemble \cite{rupp0,rupp}.

 We now briefly review the phase structures in $d=5$ as well as $d=6$ dimensions in both canonical and grand canonical ensembles.  We will identify the phase transitions and the critical phenomena in each case and describe the scaling behavior near the critical point/transition point. At the outset we clarify that we shall not be undertaking a detailed mathematical analysis of the phase behavior and wherever possible will rather explain the phase structure pictorially. For details we refer the interested reader to \cite{sudipt1,sudipt2,cai}.  In \cite{sudipt1, sudipt2} the authors offer an especially attractive Bragg-Williams formulation of black hole phase transitions.

\subsection{$d=5$ case}

For $d=5$ the expressions for $M,S$ and $T$ are simplified to the following;
\begin{align}
M&=\frac{\pi  \left(Q^2+12 r^4+12 r^6+24 r^2 \alpha \right)}{32 r^2}\label{m5d}\\
T&=\frac{-Q^2+12 r^4+24 r^6}{24 \pi  r^3 \left(r^2+4 \alpha \right)}\label{t5d}\\
S&=\frac{1}{2} \pi ^2 r \left(r^2+12 \alpha \right)\label{s5d}\\
\Phi &= \frac{\pi  Q}{16 r^2}\label{fi5d}\\
\end{align}

For completeness, we briefly sketch the uncharged and charged canonical ensembles before discussing in detail the charged grand canonical ensemble which directly connects with this work.

The uncharged case or the Schwarzschild-Gauss-Bonnet AdS black hole already shows phase coexistence and criticality in the canonical ensemble. The response coefficient relevant to this case is the heat capacity $C=T\,(\partial S/\partial T)$ and is given as
\begin{equation}
C=\frac{3 \pi ^2 r \left(1+2 r^2\right) \left(r^2+4 \alpha \right)^2}{4 r^4+8 \alpha +r^2 (-2+48 \alpha )}
\label{cuncharged}
\end{equation}
while the Helmholtz free energy $F=E-TS$ is given as
\begin{equation}
F=-\frac{\pi  \left(-r^4+r^6+6 r^2 \alpha +36 r^4 \alpha -24 \alpha ^2\right)}{8 \left(r^2+4 \alpha \right)}
\end{equation}

The locus of zeroes of the denominator of $C$ is the spinodal curve in the $r-\alpha$ plane which demarcates the region of thermally stable positive $C$ from the thermally unstable negative $C$. It is rather easier to obtain the fixed spacetime parameter $\alpha$ in terms of $r$ and plot the spinodal in the $\alpha-r_h$ plane,
\begin{equation}
\alpha=r^2\,\frac{1-2 r^2}{4 \left(1+6 r^2\right)}
\label{alphaspinodal}
\end{equation}

\begin{figure}[htp]
\includegraphics[width=0.5\textwidth, height=8cm]{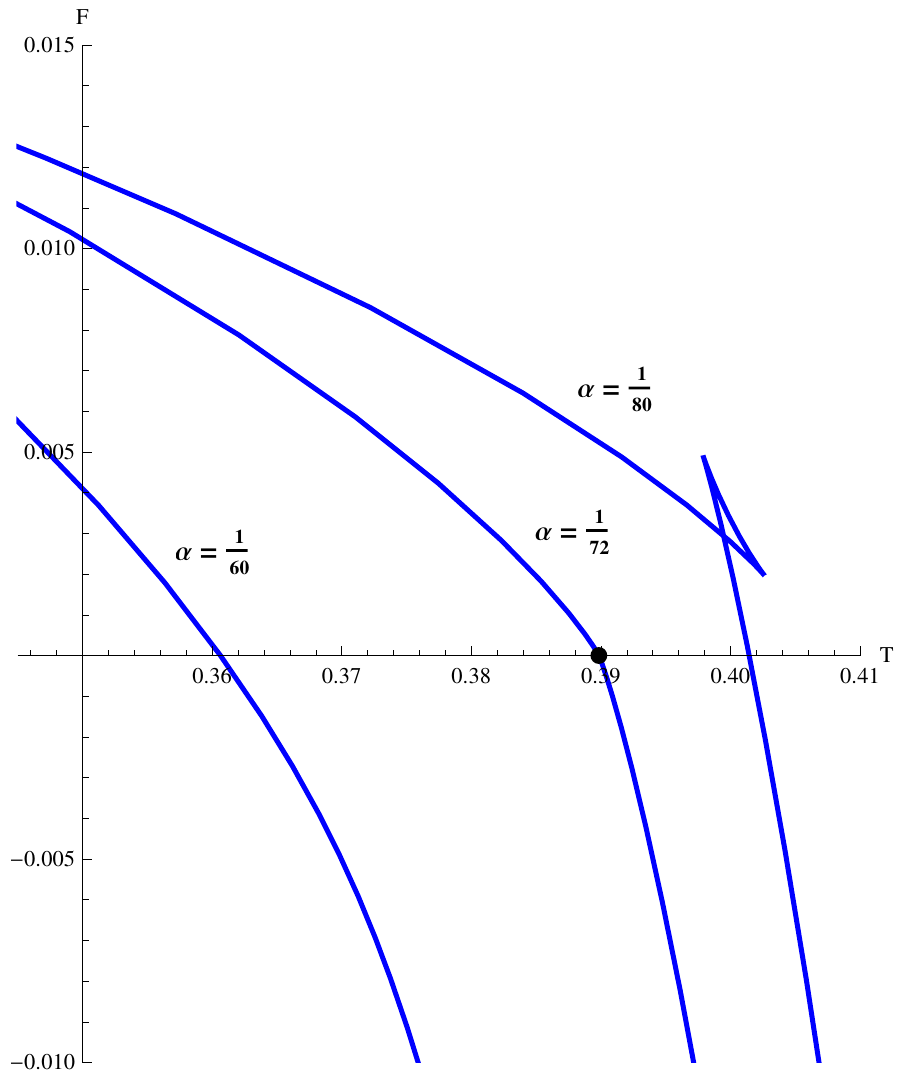}
\caption{Helmholtz free energy vs. temperature curves in the canonical ensemble for the uncharged Gauss-Bonnet AdS black hole in $d=5$. The numbers against the curves are values of $\alpha$. The critical point, labeled by a black  dot at $T_{cr}=\frac{1}{\pi}\sqrt{\frac{3}{2}}$, is coincidentally also the Hawking-Page transition point for the critical curve.}
\label{uncharged1}
\end{figure}

Similarly, the Hawking-Page curve is obtained from the zero of $F$ as,
\begin{equation}
\alpha=\frac{1}{24} \left(3 r^2+18 r^4\pm\sqrt{3} \sqrt{-5 r^4+44 r^6+108 r^8}\right)
\end{equation}
\begin{figure}[t]
\includegraphics[width=0.5\textwidth, height=6cm]{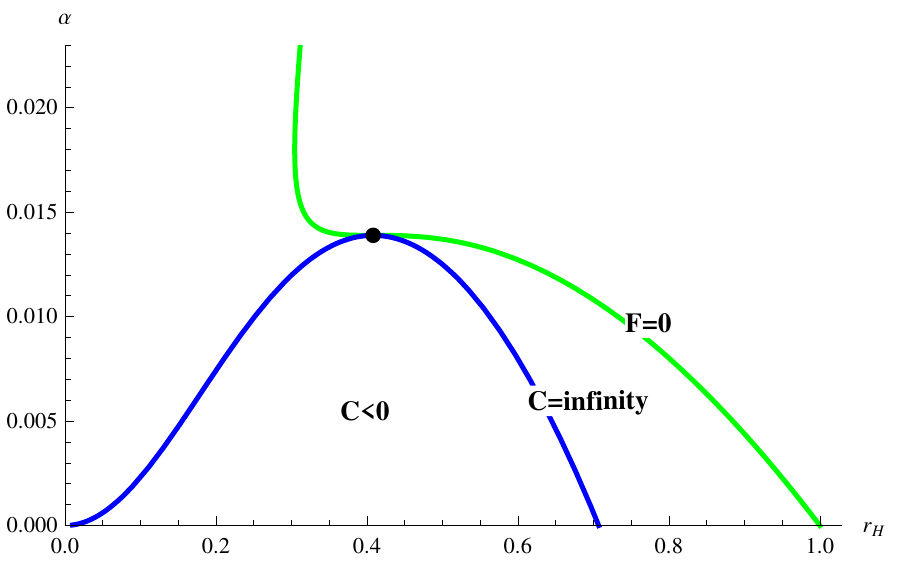}
\caption{Phase structure in the $\alpha-r_h$ plane of the canonical ensemble for $d=5$ uncharged Gauss-Bonnet-=AdS black holes. The blue curve is akin to a spinodal along which the heat capacity $C$ diverges and is negative under it. The green curve is the Hawking-Page curve. The critical point, labeled by a black  dot, is at $\alpha=1/72$, $r_h=1/\sqrt[]{6}$.}
\label{uncharged}
\end{figure}
In figure (\ref{uncharged}), we plot the blue coded spinodal and the green coded Hawking-Page curve in the $\alpha-r_h$ plane. On heating the black hole at any fixed $0<\alpha<1/72$ it undergoes a first order phase transition from the small black hole phase (which starts at $T=0$ with $r_h =0$) to the large black hole phase via a phase coexistence region. Curiously, the critical point at $\alpha=1/72$ is also the Hawking-Page transition point. Note that in figure (\ref{uncharged}), moving on horizontal lines where an $\alpha$ is fixed is only meaningful. In figure (\ref{uncharged1}), we plot the free energy versus temperature for sub-critical, critical and super-critical values of alpha. In the sub-critical region, the swallow tail curve of the free energy gives the first order phase transition point. Note that since we are not considering ensembles in which $\alpha$ might fluctuate, hence each value of $\alpha$ represents a fixed spacetime and we are allowed only to have those paths on which $\alpha$ remains fixed.

\subsubsection{Canonical Ensemble}

In the canonical ensemble of the charged case the black hole charge remains fixed so that the only means of exchanging energy with the surroundings is via heat transfer. This reduces the first law  to 
\begin{equation}
dS=\frac{1}{T}\,dM_Q
\label{canonical law}
\end{equation}
where, just for this equation, we have sneaked in the subscript $Q$ to emphasize that the fixed, non-fluctuating charge of the black hole is not a thermodynamic variable but a parameter in the canonical ensemble. Of course $Q=0$ reduces to the uncharged case discussed earlier.

The response coefficient relevant to this case is the heat capacity at fixed charge, 

\begin{align*}
& C_Q =T\left(\frac{\partial M}{\partial T}\right)_Q\\
&=\frac{3 \pi ^2 r \left(-Q^2+12 r^4+24 r^6\right) \left(r^2+4 \alpha \right)^2}{2 \left(5 Q^2 r^2-12 r^6+24 r^8+12 Q^2 \alpha +48 r^4 \alpha +288 r^6 \alpha \right)}\label{cq5d}\numberthis
\end{align*}

\begin{figure}
\includegraphics[width=0.5\textwidth, height=6cm]{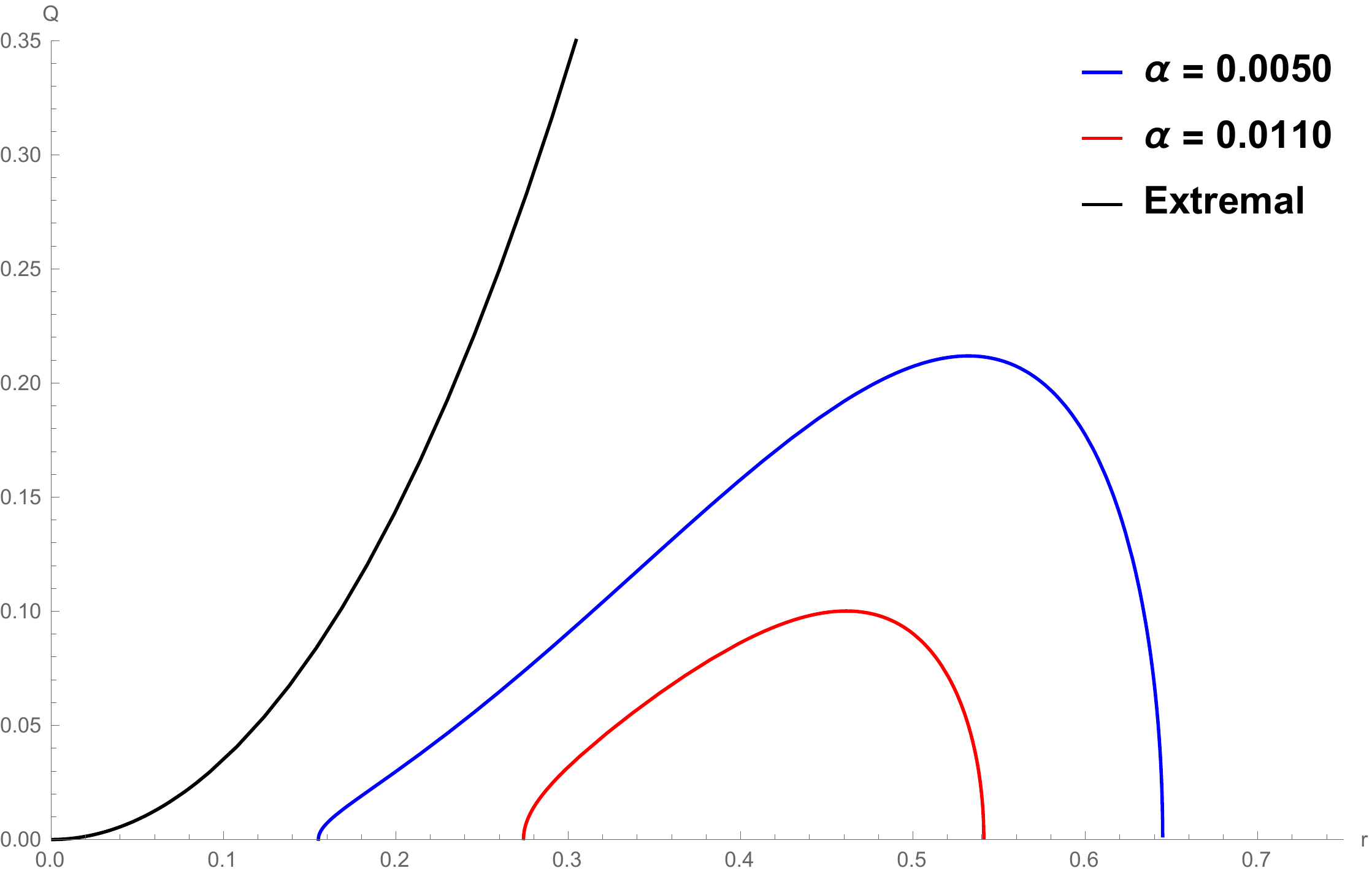}
\caption{Phase structure in the $Q-r_h$ plane of the  $d = 5$ GB-AdS black holes in the fixed charge ensemble showing the spinodals for $\alpha = 0.0050$ (blue) and $\alpha = 0.0110$ (red). Black line shows the extremal curve left to which is $T<0$. The spinodal shrinks and disappears at $\alpha=1/72$. For higher $\alpha$ small black holes become large black holes without any phase transition. }
\label{phasecanon}
\end{figure}

The heat capacity changes sign through an infinite dicontinuity at the zero of its denominator in eq. (\ref{cq5d}) above so that the spinodal in the $Q-r_h$ plane for a fixed $\alpha$ is given as
\begin{equation}
Q^2_{spinodal}=\frac{12 \left(r^6-2 r^8-4 r^4 \alpha -24 r^6 \alpha \right)}{5 r^2+12 \alpha }
\label{cqdiv}
\end{equation}
as shown in the phase structure plot of figure (\ref{phasecanon}). The heat capacity is negative inside the spinodal and positive outside. Therefore, a line of constant $Q$ in figure (\ref{phasecanon}) would typically pass from the locally stable small black hole (SBH) phase to the locally stable large black hole phase (LBH) via the unstable branch. The line tangent to the locus would hit the critical point in passing from the SBH to the LBH phase. The critical horizon radius is easily found as the maximum of the spinodal in eq. (\ref{cqdiv}) above as 
 \begin{equation}
 r^2_{cr}=\frac{1}{30} \left(5-108 \alpha +\sqrt{25-1800 \alpha +11664 \alpha ^2}\right)
 \label{rcritcan}
 \end{equation}
 
 The critical charge for a given $\alpha$ can be found by substituting eq. (\ref{rcritcan}) in eq. (\ref{cqdiv}) above. One may check that the critical charge thus obtained becomes zero for $\alpha=1/72$. Therefore, just as for the uncharged case, phase coexistence and criticality ceases for $\alpha>1/72$. In figure (\ref{phasecanon}), this is reflected in the observation that the spinodal shrinks to zero as $\alpha$ increases to $\alpha=1/72$. We add that the black hole is globally unstable against charged, thermal GB-AdS in the regions of phase coexistence as can be checked by calculating its Helmholtz free energy $F$
 
  
   We should emphasize here that because of the nature of the canonical ensemble, it is only possible for a given black hole to move along a line of constant charge, whether in the $S-T$ plane or the $Q-R$ plane. In particular, for example, it is not possible to move along an isotherm without varying the charge and thus breaking away from the ensemble. Moreover, since the black hole cannot vary its charge here, it can only hit criticality if it is already prepared with a critical charge for a given $\alpha$. Similarly, if it is prepared with a sub-critical charge the black hole is bound to undergo a first order phase transition. More directly, there is only one dimension for any maneuverability in the thermodynamic state space, as implied by eq. (\ref{canonical law}) above. It is essentially this constraint that has resulted in attempts to obtain scaling variables near criticality that are not fully consistent. These attempts have been nicely reviewed and critiqued in \citep{mann}. The same reference also discusses $P-V$ criticality in the extended state space of the RN-AdS black hole in the canonical ensemble. Indeed, allowing for AdS length scale as a variable gets the necessary purchase for a genuine criticality in the canonical ensemble by adding another dimension to  the thermodynamic state space. Thus, in particular, one can now move along curves of ${\it both}$ fixed $Q$ and $T$ in the extended state space by varying the pressure. This results in the well known $P-V$ curves for the black holes which are exactly analogous to the ones for the van der Waals gas. Moreover, with the pressure and the temperature as the scaling variables one finds the critical exponents in the standard manner and these are shown to have mean field values. 
   
   Unfortunately for the Ruppeiner geometry though, the extended state space is still unable to rescue the canonical ensemble. This is because, as mentioned in the preceding, for charged static AdS black holes in general, the fluctuations of the thermodynamic volume are not  independent of entropy and hence the {\it manifold of fluctuations} still remains one dimensional.

\subsubsection{Grand Canonical Ensemble}
  
It turns out for the GB-AdS black holes, the grand canonical ensemble with both $M$ and $Q$ fluctuations still displays phase coexistence and criticality. It is indeed a rare occurrence so  far as black holes in non-extended phase space are concerned and renders it particularly amenable to a TG analysis. Earlier work in which TG could address criticality in the non-extended black holes was in the mixed ensembles of Kerr-Newman-AdS (KN-AdS) black holes, wherein either the charge or the angular momentum were held fixed while the other quantity was allowed to vary (and hence spontaneously fluctuate) thus ensuring a two-dimensional thermodynamic manifold \citep{sahay2,sahay3}.

  In the grand canonical ensemble the black hole is free to  exchange its charge and mass with the surrounding which is held at a fixed temperature and potential. The Euclidean action eq. (\ref{eucidean}), and hence the Gibbs function with its fixed gauge potential quite naturally describes the grand ensemble. The Gibbs free energy is obtained as follows,
  \begin{widetext}
\begin{equation}
G=-\frac{3 \pi ^2 \left(r^6+6 r^2 \alpha -24 \alpha ^2-r^4 (1-36 \alpha )\right)+64 r^2 \left(r^2-12 \alpha \right) \Phi ^2}{24 \pi  \left(r^2+4 \alpha \right)}
\label{g5d}
\end{equation}
\end{widetext} 

The zero of $G$ determines the Hawking-Page transition point between the charged GB-AdS black hole and thermal AdS. Note that it is convenient to replace $Q$ by $\Phi$ in the expressions for thermodynamic variables in the grand canonical ensemble as we have done in the equation above. Thus, for example, the temperature for the $5-d$ case can be written as
\begin{equation}
T=\frac{r \left(\pi ^2 \left(3+6 r^2\right)-64 \Phi ^2\right)}{6 \pi ^3 \left(r^2+4 \alpha \right)}
\label{tphi5d}
\end{equation}

The zero of the temperature is controlled by the following function,

\begin{equation}
\mathcal{Z}_5=\pi ^2 \left(3+6 r^2\right)-64 \Phi ^2
\label{z5}
\end{equation}

The three second partial derivatives of $G$ with respect to $T$ and $\Phi$ give rise to three independent response functions relevant to this ensemble. They are the heat capacity $C_\Phi$, the isothermal capacitance and the thermal `expansivity' of the charge, given in order as
\begin{widetext}
\begin{align}
T\left(\frac{\partial S}{\partial T}\right)_{\Phi}&=\frac{3 \pi ^2 r \left(r^2+4 \alpha \right)^2 \left(\pi ^2 \left(3+6 r^2\right)-64 \Phi ^2\right)}{2 \left(3 \pi ^2 \left(2 r^4+4 \alpha -r^2 (1-24 \alpha )\right)+64 \left(r^2-4 \alpha \right) \Phi ^2\right)}\label{cphi5}\\ 
\left(\frac{\partial Q}{\partial \Phi}\right)_{T}
&=\frac{16 r^2 \left(3 \pi ^2 \left(2 r^4+4 \alpha -r^2 (1-24 \alpha )\right)+64 \left(5 r^2+12 \alpha \right) \Phi ^2\right)}{\pi  \left(3 \pi ^2 \left(2 r^4+4 \alpha -r^2 (1-24 \alpha )\right)+64 \left(r^2-4 \alpha \right) \Phi ^2\right)}\label{cap5}\\
\left(\frac{\partial Q}{\partial T}\right)_{\Phi}&=\frac{192 \pi ^2 r \left(r^2+4 \alpha \right)^2 \Phi }{3 \pi ^2 \left(2 r^4+4 \alpha -r^2 (1-24 \alpha )\right)+64 \left(r^2-4 \alpha \right) \Phi ^2}\label{exp5}
\end{align}
\end{widetext}

The divergence of the response functions is  seen to be controlled by the following polyomial in $r$,
\begin{align}
\mathcal{N}_5=3 \pi ^2 \left(2 r^4+4 \alpha -r^2 (1-24 \alpha )\right)+64 \left(r^2-4 \alpha \right) \Phi ^2
\label{n5}
\end{align}

\begin{figure}[htp]
\includegraphics[width=0.5\textwidth, height=6cm]{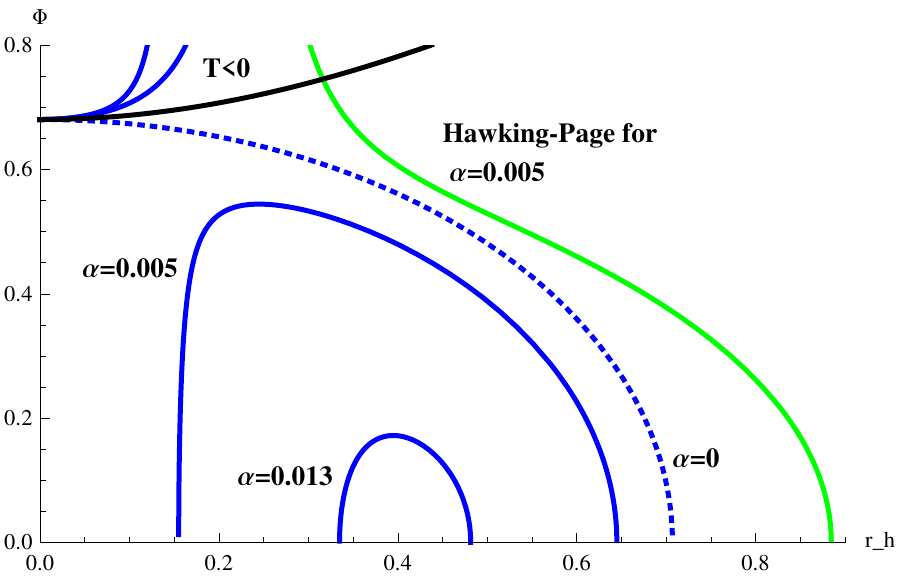}
\caption{Phase structure in the $\Phi-r_h$ plane of $d=5$ GB-AdS black holes in the grand canonical ensemble, with the spinodal curves for $\alpha=0.0130$ (blue) and $\alpha=0.0050$ (blue) and, for comparison, $\alpha=0$ (blue dotted). The black coded extremal curve is independent of $\alpha$ in $d=5$. The green coded curve is the Hawking-Page curve for $\alpha=0.0050$ so that the Gibbs energy of the black hole is negative to the right of the curve.}
\label{phasegrand5}
\end{figure}

In figure (\ref{phasegrand5}) we plot a graph of the spinodal curves or the locus of zeroes of $\mathcal{N}_5$ , zeroes of $T$ and the zeroes of $G$ in the $\Phi-r$ plane. For comparison, we show  spinodals at three different $\alpha$'s including the $5-d$ RN-AdS case with $\alpha=0$. It is interesting to note the qualitative change in the phase structure for any non-zero $\alpha$, however small. In a way, therefore, even a very weak Gauss-Bonnet correction to the Einstein-Hilbert theory effects a non-trivial change in the phase dynamics of the black hole. It can be checked that for $\alpha>1/72=0.0139$ the spinodal curve shrinks to zero so that phase coexistence disappears and there is a thermodynamic continuity between the small and large black holes marked by the absence of singularity in the response functions. Nevertheless, analogous to the supercritical region for ordinary systems, the {\it anomalies} in response persist for a while. We shall have more to speak about the parameter space $\alpha>1/72$ shortly. Coming back to figure (\ref{phasegrand5}), three different developments of phase structure can be seen depending on the range of values of $\Phi$. Starting from $\Phi=0$, the black hole goes through a phase coexistence region much like the canonical ensemble. However, note a key difference. The small black hole starts all the way from $r=0$ unlike the canonical case and hence there is no extremality. There is a critical point at 
\begin{align*}
\Phi_{cr}&=\frac{\sqrt{3}\pi}{8}\sqrt{1-72\alpha}\\Q_{cr}&=24 \sqrt{3} \alpha \sqrt{1-72 \alpha }\\T_{cr}&=\frac{6}{\pi}\sqrt{3 \alpha}
\label{rcr5}\numberthis
\end{align*}
and $r_{cr}=2 \sqrt{3\alpha }$, where we have used eq. (\ref{n5}) to obtain the critical point. In the supercritical, single phase region, for $\Phi_{cr} < \Phi <\Phi_0=\sqrt{3}\pi/8$ there are still no extremal black holes. For $\Phi > \Phi_0$ extremal black holes appear. Note that the entire phase coexistence region in figure (\ref{phasegrand5}) lies to the left of the Hawking-Page curve and hence is globally unstable against thermal Gauss-Bonnet AdS at a fixed pure gauge potential.

\begin{figure}
\includegraphics[width=0.5\textwidth, height=7cm]{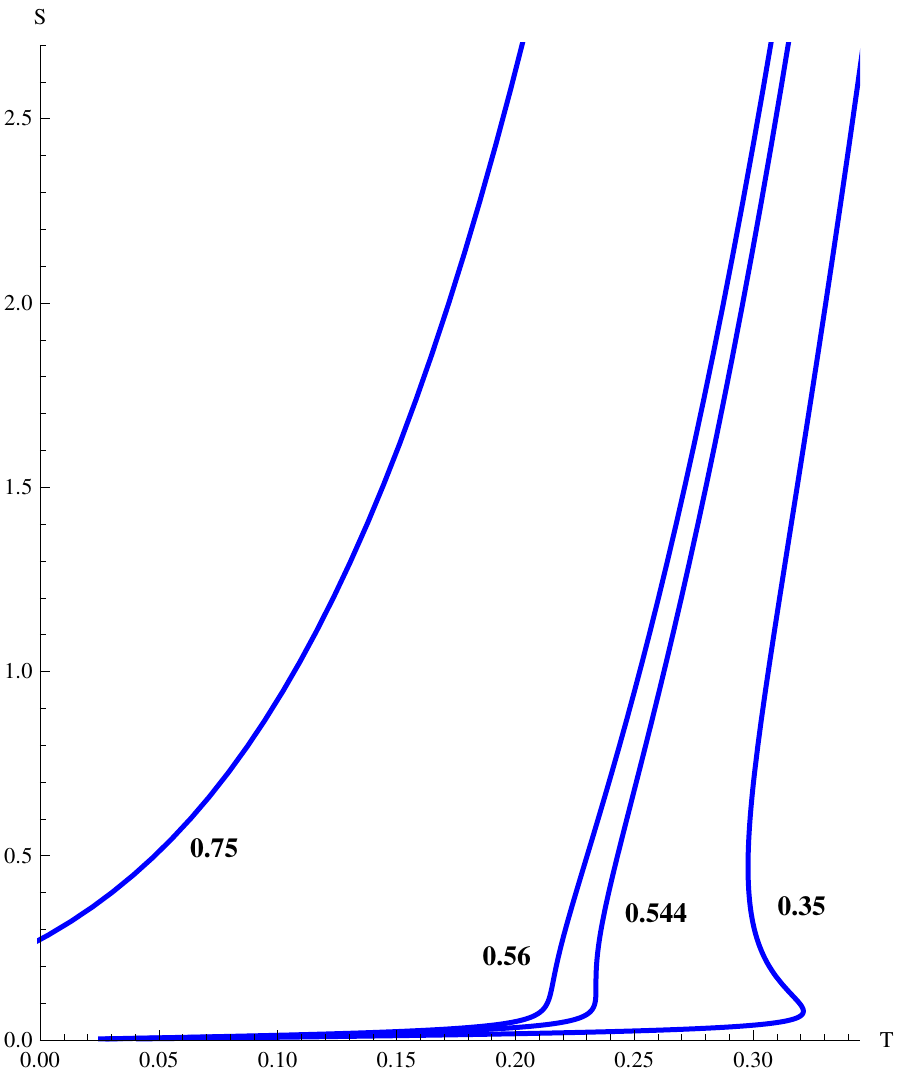}
\caption{Isopotential plots for representative values of potential in $5-d$ GB-AdS black holes with $\alpha=0.0050$. The critical potential for $\alpha=0.0050$ is $\Phi_{cr}=0.5440$. For higher values of potential extremal black holes appear.}
\label{granst}
\end{figure}

 In figure (\ref{granst}), we plot isopotenial $S-T$ curves with representative potential values for each region mentioned above. Note that curve at $\Phi=0.56$ which is close to the critical isopotential at $\Phi=0.544$ shows a sharp change from the small to the large black hole phase.  This is therefore akin to a supercritical phase, a typical feature of mean field models with an isolated critical point like the van der Waals gas. Of course, for ordinary thermodynamic systems, of which such mean field models are approximate representations, supercriticality is a matter of fact. Thus, the line of extrema of correlation length or the Widom line is thought to uniquely mark the cross-over from one phase to the other in the supercritical region close to the critical point \cite{stan,scop}. An important line of investigation in this work is to see whether there exists a unique characterization of supercriticality for black holes. As we shall discuss in a subsequent section, geometry seems to suggest a tentative answer in the affirmative.

\begin{figure}[htp]
\includegraphics[width=0.5\textwidth, height=6cm]{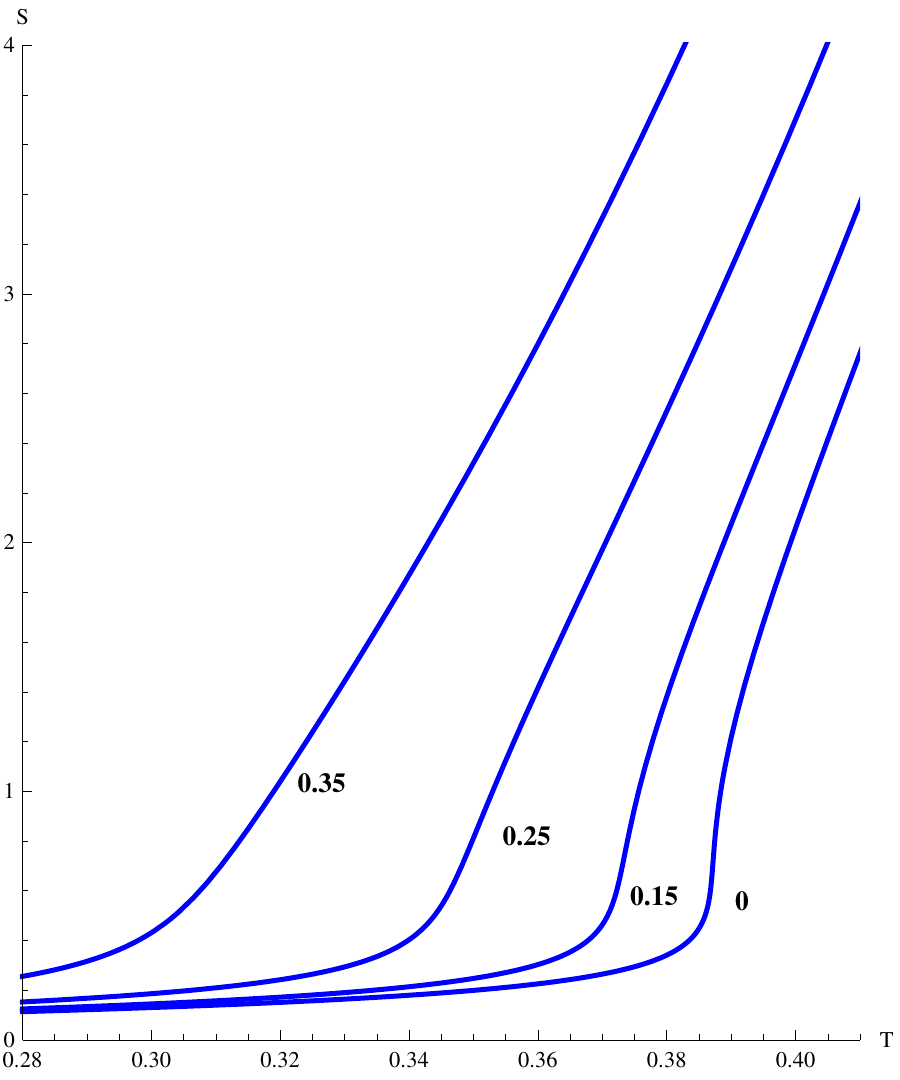}
\caption{Isopotential plots in the $S-T$ plane of $d=5$ GB-AdS black holes in the grand canonical ensemble for a $\alpha=1/70>1/72$. The potential values are indicated against the plots. }
\label{superalpha}
\end{figure}

In figure (\ref{superalpha}), we plot isopotential curves in the $S-T$ plane for a value of $\alpha$ greater than $1/72$. At the lowest potential namely $\Phi=0$ the isopotential curve looks very much like a supercritical curve close to criticality. As $\Phi$ increases this feature gets more and more diminished until at higher potentials the curve becomes featureless. We remind ourselves that the zero potential limit in the grand canonical ensemble is not equivalent to the Schwarzschild-GB-AdS limit since in the former case charge fluctuations remain finite at zero potential.

   The grand canonical ensemble is less constrained as compared to the canonical ensemble and, as a result, its region of local stability shrinks compared to the canonical case \cite{sahay}. While this is true it does not imply that a given black  hole in the canonical ensemble has access to more stable regions. In fact, it is quite the opposite. With an additional degree of freedom in the thermodynamic state space afforded by the grand ensemble vis-a-vis the canonical  ensemble the black hole has now many mores thermodynamic states accessible to it. In particular, it can approach the critical point in any direction. Contrast this with the canonical case where, in the non-extended case, evolving the black hole along an isotherm already violates the fixed charge condition.
  

 \subsubsection{Scaling near the $\Phi-Q$ critical point}
 
At the critical  point in the grand canonical ensemble the black hole undergoes a continuous phase transition from the small black hole phase to the large black hole phase. The scaling behavior of all the thermodynamic functions near the critical point can be found by observing the scaling of the singular part of the Gibbs free energy. 

The scaling of the thermodynamic quantities can be suitably ascertained by a consideration of the behaviour near the critical point of the following equation of state giving a relation between $T,Q$ and $\Phi$
 
\begin{equation}
T=\frac{\sqrt{Q} \left(3 \pi ^3 Q+24 \pi ^2 \Phi -512 \Phi ^3\right)}{12 \pi ^{5/2} \sqrt{\Phi } (\pi  Q+64 \alpha  \Phi )}
\label{eos5}
\end{equation}

The above equation can be obtained by solving eq. (\ref{phi}) in $5-d$ for $r$ and then plugging it into eq. (\ref{tphi5d}) for $T$. To fix an analogy with the liquid-vapor critical phenomena, the order parameter $Q$ (more precisely $Q_{lbh}-Q_{sbh}$ or $Q-Q_{cr}$) is the counterpart of the fluid volume while the conjugate `ordering field' $\Phi$ is like the hydrodynamic pressure. It is not possible to invert the equation above and obtain the standard equation of state form, $\Phi(Q,T)$. Nevertheless, in the neighborhood of the critical point given by eq.(\ref{rcr5}), the critical equation of state can be obtained in the standard form,

\begin{equation}
{\Phi}=\frac{3\alpha}{1-72\alpha}\left(\frac{1}{2}{q}^3-16{t}-4{q}{t}\right)+ \ldots\label{criteos}
\end{equation} 

where the deviations from criticality are given by ${\Phi}=(\Phi_{cr}-\Phi)/\Phi_{cr}$, ${t}=(T-T_{cr})/T_{cr}$, and ${q}=(Q-Q_{cr})/Q_{cr}.$ The limiting slope of the coexistence curve in the $\Phi-T$ plane is readily observed from the critical equation of state.
\begin{equation}
\left.\frac{dP}{dT}\right|_{sat}^{cr}=\left(\frac{\partial \Phi}{\partial T}\right)_{Q_{cr}}=-\frac{48\alpha}{1-72\alpha}
\label{asymcoex}
\end{equation}

 The $\Phi-T$ coexistence curve slopes to the left as opposed to the $P-T$ coexistence curve for the van der Waals fluid and is reminiscent of the anomalous water-ice curve.

The exponent for the degree of the critical isotherm $t=0$ is easily seen to be $\delta=3$ by setting $t=0$ in (\ref{criteos}) above, 
\begin{equation}
\Phi=\frac{3 \alpha}{2(1-72\alpha)}q^3
\end{equation}
The critical exponent  $\beta$, which is the degree of the $T-Q$ coexistence curve near crticality, can now be obtained by a standard procedure via the Maxwell construction as outlined in \cite{ mann, zou}. In any case it is easy to see from eq. (\ref{asymcoex}) above that, with the reasonable assumption that $\Phi$ has a linear growth with $t$ in the asymptotic critical region, the middle term in the $r.h.s$ of eq.(\ref{criteos}) above cancels the $\Phi$ in its $l.h.s$ thereby giving,

\begin{equation}
q_{sbh}=-q_{lbh}=\pm\, 2\sqrt {2}\,\sqrt{t}
\label{beta expo}
\end{equation}
along the coexistence curve in the critical region. Thus one finds that $\beta=1/2$.

The exponent $\gamma$ gives the scaling  of the relevant susceptibility with respect to ${t}$ on approaching the critical point along a critical isochore. In this case, therefore, the scaling of the capacitance $(\frac{\partial Q}{\partial \Phi})_T $ has to be observed along the isocharge $Q=Q_{cr}$. Again, from equation (\ref{criteos}) above one finds that 
\begin{equation}
\left.\frac{\partial {\Phi}}{\partial {q}}\right|_{t}=\frac{3} {32}\,\tilde{t}
\label{gama expo}
\end{equation}
 along the isocharge ${q}=0$ in the critical region thus giving $\gamma=1$. Lastly, $\alpha$ (not to be confused with the GB coupling of the same name) which gives the scaling of the heat capacity at fixed order parameter, refers to the critical behaviour of $C_Q$ in this case along the critical isopotential $\Phi=\Phi_{cr}$. Since the heat capacity at constant charge (eq. (\ref{cq5d})) is quite regular at the critical point pertaining to the grand canonical ensemble we have $\alpha=0$. 
 
 Indeed, the standard scaling exponents for the $\Phi-Q$ criticality take mean field values,
\begin{equation}
\alpha=0\,\,,\,\,\beta=\frac{1}{2}\,\,,\,\,\delta=3\,\,,\,\,\gamma=1\label{expo}
\end{equation}

Let us now consider the scaling of the Gibbs free energy, eq. (\ref{g5d}) as we approach the critical point from the supercritical region along the critical isotherm and the critical isopot. We chose these two paths with a view to a comparison with the scaling behaviour of the thermodynamic curvature $R$ in a subsequent section.  A direct way to obtain the scaling is to first invert eq. (\ref{tphi5d}) above to obtain $r$ in terms of $T$ and $\Phi$. This is possible in spite of a third degree equation because in the supercritical region $r$ is monotonic function of both the variables. Now we plug this result in eq. (\ref{g5d}) to obtain $G$ as a function of $T$ and $\Phi$. It is then straightforward to obtain a series expansion of $G$ along, successively, the critical isotherm 
\begin{align*}
&G=\frac{3 \pi}{4} ( \alpha  (1-72 \alpha ))\\&-24 \sqrt{3} ( \alpha \sqrt{1-72 \alpha}) (\Phi- \Phi_{cr})\\&-36 \left(\frac{2}{\pi }\right)^{1/3}( \alpha^{2/3}(1 -72 \alpha)^{2/3}) (\Phi- \Phi_{cr})^{4/3}+\ldots\label{gstherm}\numberthis
\end{align*}

and the critical isopotential,

\begin{align*}
&G=\frac{3 \pi}{4} (  \alpha  (1-72 \alpha ))\\&-24\sqrt{3} \pi^2 \alpha^{3/2} \left(T-T_{\text{cr}}\right)\\&- 36 \left(2^{1/3}\right) \pi^{7/3} \alpha^{4/3} \left(T-T_{\text{cr}}\right){}^{4/3}+\ldots\label{gspot}\numberthis
\end{align*}

In both the equations above we take the dominant, singular part $G_s$ to be the third term in the series. In both cases the singular part scales as the $4/3$-rd power of $(T-T_{\text{cr}})$ and $(\Phi- \Phi_{cr})$. This is expected since the $\Phi-Q$ critical point is an inflection point. One obtains exactly the same scaling in the supercritical region for the van der Waals case with appropriate scaling variables. Note that we have displayed the amplitudes explicitly so that we may readily compare with the corresponding scaling of $R$ in a subsequent section.

\subsection{$d=6$ case}

As we have mentioned previously, the thermodynamics and phase structure of the $d=6$ case is representative of all $d\geq 6$ black holes \cite{cai,caicai}.
   
The basic thermodynamic expressions for $d=6$ are obtained as,
\begin{align}
M&=\frac{\pi  \left(Q^2 + 24 r^{4}\left(r^4 + r^2+6 \alpha \right) \right)}{36 r^{3}} \label{m6d}\\
T&=\frac{-Q^2+40 r^8+24 r^6+48 r^4 \alpha }{32 \pi  r^5(r^2+12 \alpha) } \label{t6d}\\
S&=\frac{2}{3} \pi ^2 r^2 \left(r^2 + 24 \alpha \right)\label{s6d} \\
\Phi &= \frac{\pi  Q}{18 r^3}\label{fi6d}
\end{align}

Note that, unlike the $d=5$ case, now the extremal temperature depends on $\alpha$. 

\subsubsection{Canonical Ensemble}

\begin{figure}
\includegraphics[width=0.5\textwidth, height=6cm]{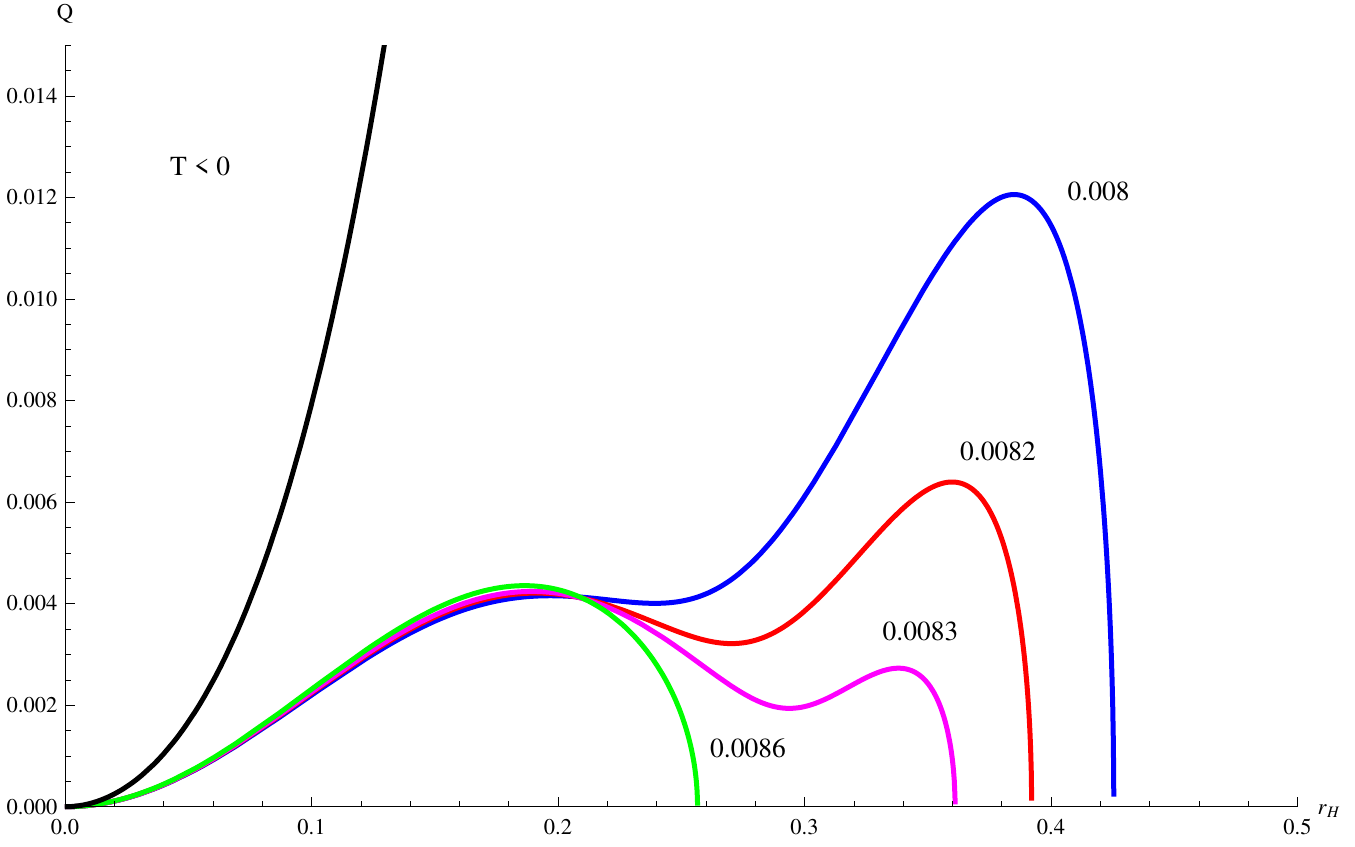}
\caption{Phase structure in the $Q-r_h$ plane of $d=6$ GB-AdS black holes for the fixed charge ensemble. The spinodals are the locus of singularity of $C_Q$ and are labeled by different values of $\alpha$. Heat capacity is negative below each spinodal. The black line curving upwards to the left is representative of the four extremal curves ($T$=0) which are very close to each other because of the proximity of the corresponding $\alpha$ values. }
\label{phasecan6}
\end{figure}

\begin{figure}
\includegraphics[width=0.5\textwidth, height=6cm]{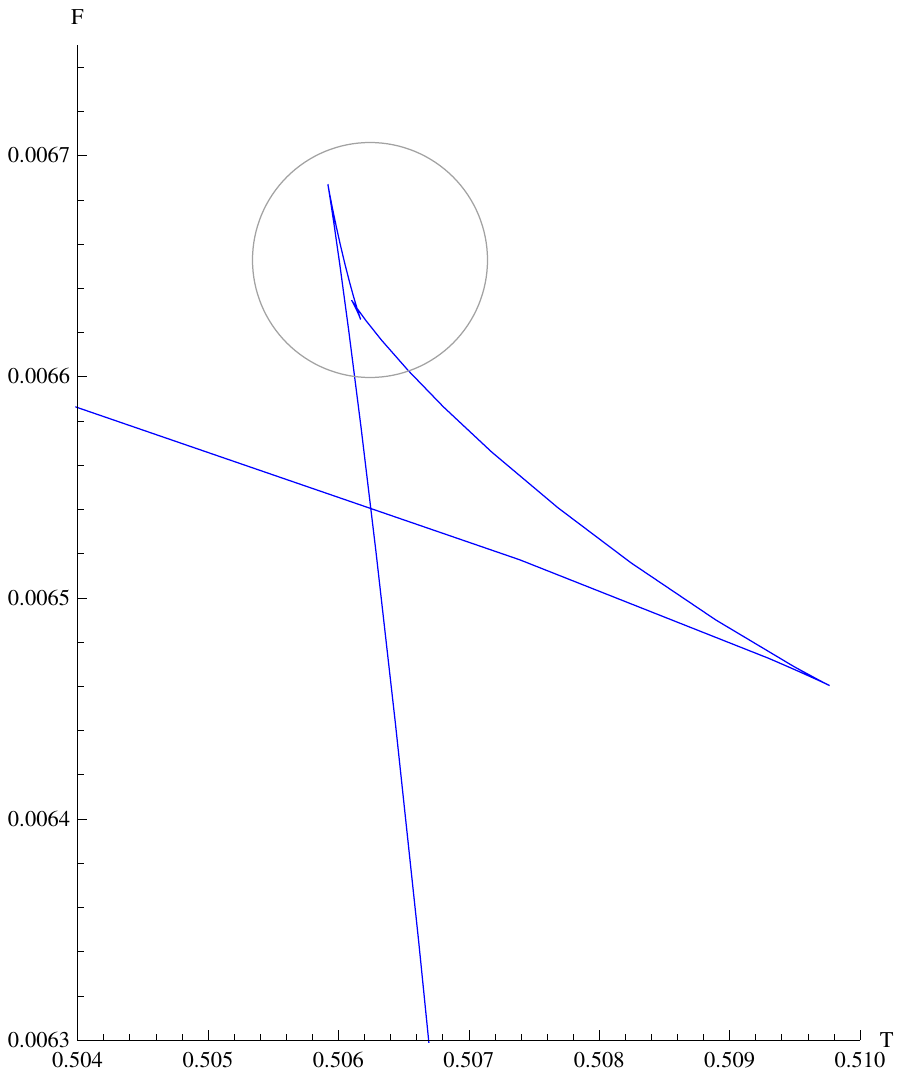}
\caption{Helmholtz free energy vs. temperature plot in the canonical ensemble of $d=6$ GB-AdS black hole with $\alpha=0.0082$ and $Q=0.00355$. The swallow tail has a window of stability in the unstable branch. Figure (\ref{d2}) is a blow up of the encircled region. }
\label{d1}
\end{figure}

\begin{figure}
\includegraphics[width=0.5\textwidth, height=6cm]{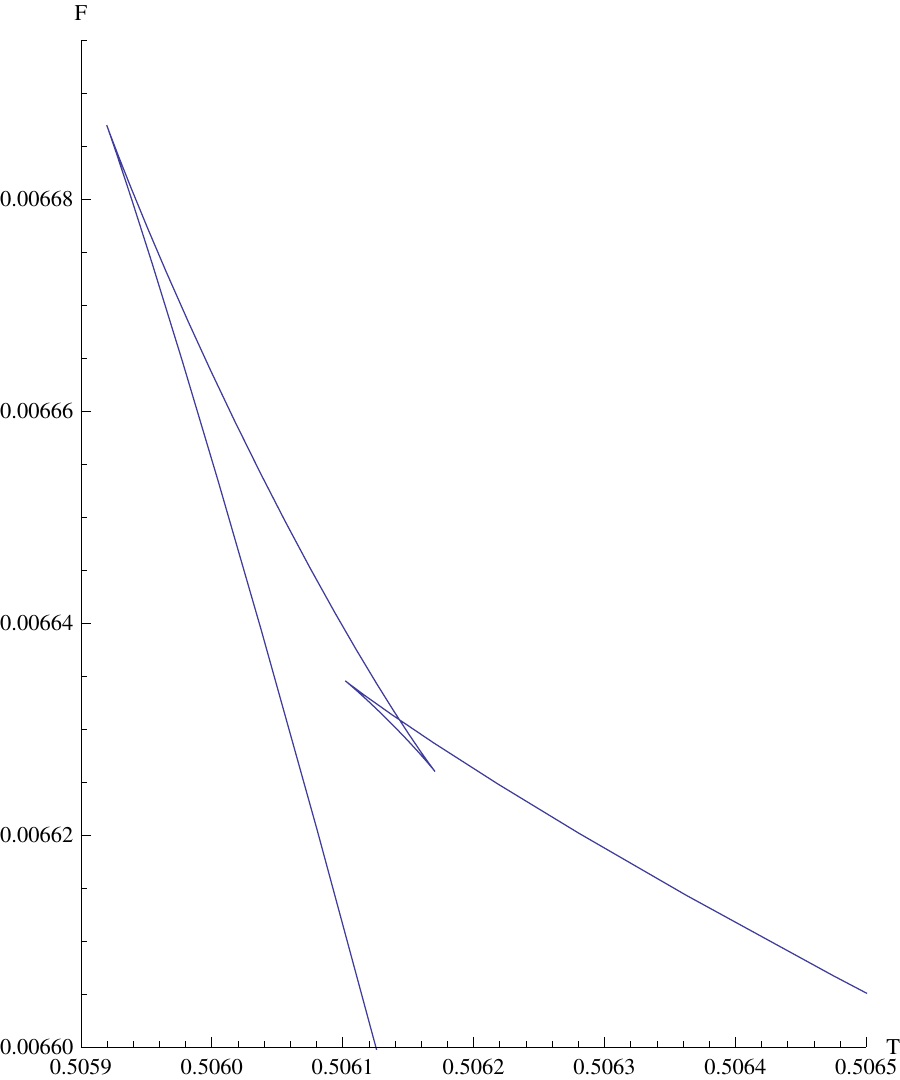}
\caption{A magnified plot of the encircled region in figure (\ref{d1}). The second swallow tail representing a window of stability in the unstable branch is clearly visible.}

\label{d2}
\end{figure}

\begin{figure}
\includegraphics[width=0.5\textwidth, height=6cm]{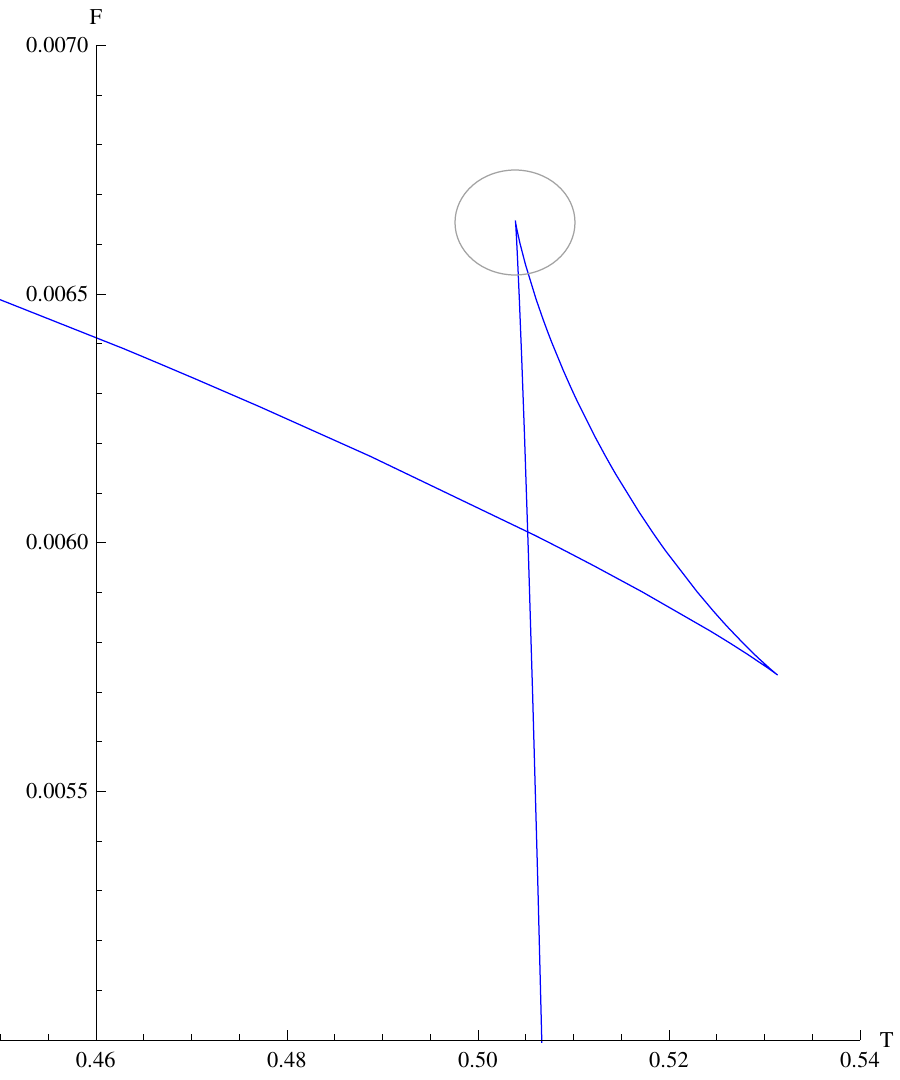}
\caption{Helmholtz free energy vs. temperature plot in the canonical ensemble of $d=6$ GB-AdS black hole with $\alpha=0.0083$ and $Q=0.00240$. The metastable part of the large black hole branch develops a further swallow tail. Figure (\ref{d12}) is a blow up of the encircled region. }
\label{d11}
\end{figure}

Let us first briefly review the canonical ensemble. The response coefficient relevant to the ensemble is $C_Q$ which is as follows,
\begin{widetext}
\begin{equation}
C_Q=\frac{8 \pi ^2 r^2 \left(r^2+12 \alpha \right)^2 \left(-Q^2+24 r^6+40 r^8+48 r^4 \alpha \right)}{3 \left(7 Q^2 r^2-24 r^8+40 r^{10}+60 Q^2 \alpha +144 r^6 \alpha +1440 r^8 \alpha -576 r^4 \alpha ^2\right)}\label{cq6}
\end{equation}
\end{widetext}

The spinodal equation is conveniently obtained by setting to zero the denominator of $C_Q$,
\begin{equation}
Q^2_{\mbox{spinodal}}=\frac{8 r^4 \left( 72 \alpha ^2-5 r^6-18 r^2 \alpha -3 r^4 (-1+60 \alpha )\right)}{7 r^2+60 \alpha }.
\label{qspincan6}
\end{equation}

The Helmholtz free energy $F=E-TS$ becomes
\begin{widetext}
\begin{equation}
F= \frac{\pi  \left(Q^2 \left(7 r^2+120 \alpha \right)-24 r^4 \left(r^6+6 r^2 \alpha -144 \alpha ^2+r^4 (-1+72 \alpha )\right)\right)}{144 r^3 \left(r^2+12 \alpha \right)}.
\label{free6dcan}
\end{equation}
\end{widetext}

In figure (\ref{phasecan6}), we plot the phase structure in the $Q-r_H$ plane based on the same guidelines as the canonical ensemble for the $d=5$ case. 


 The phase structure for the canonical ensemble of the $d=6$ case is roughly similar to the $d=5$ canonical case. However there is an interesting window of $\alpha$ values for which there is an intervening region of local stability between two regions of instability. In figure (\ref{phasecan6}), we draw a few spinodals in precisely this window. As $\alpha$ increases from zero ($6-d$ RN-AdS case) the left side of the spinodal develops an inflection at $(r, \alpha,Q,T)=(0.2121, 0.0079, 0.0041, 0.5138)$. On further increasing $\alpha$ the `hump' on the left side grows and soon dominates the one on the right, while the latter on flattens and disappears after passing through another inflection at $(r, \alpha,Q,T)=(0.3162, 0.0083, 0.0002, 0.5040)$.
 
\begin{figure}
\includegraphics[width=0.5\textwidth, height=6cm]{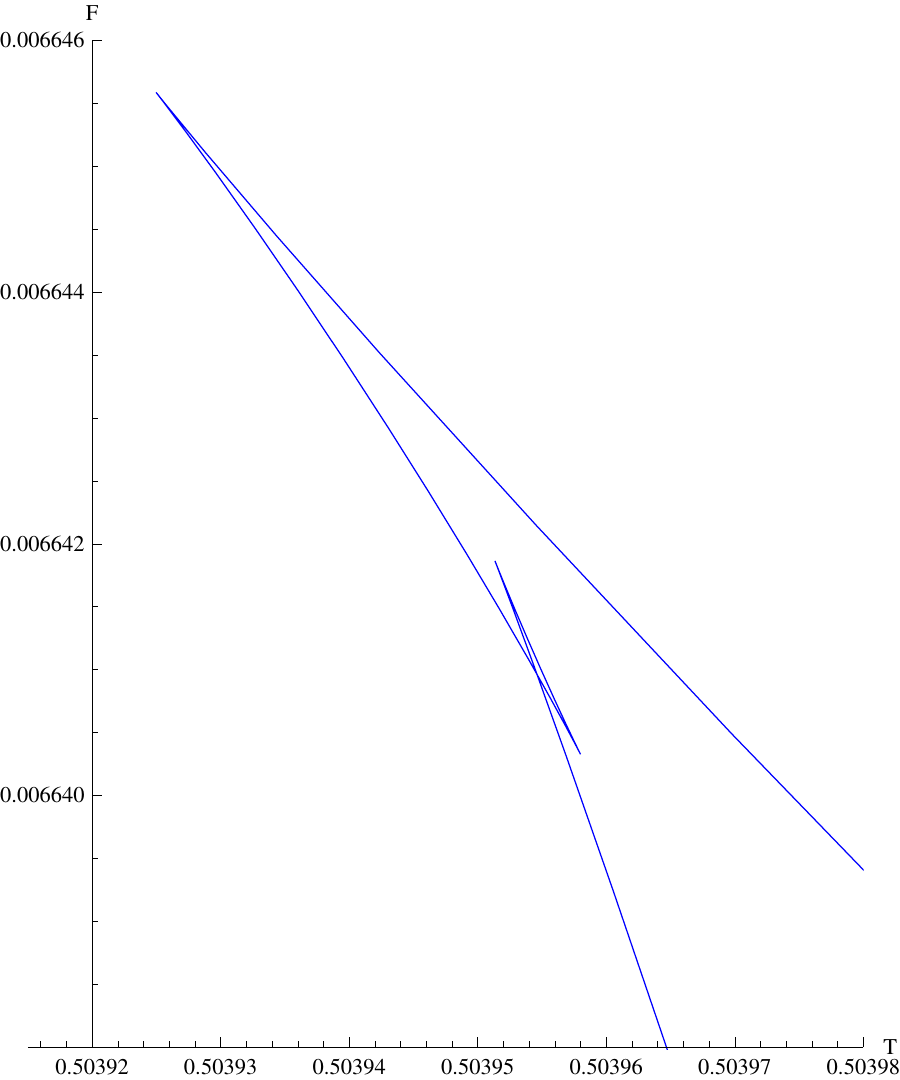}
\caption{A magnified plot of the encircled region in figure (\ref{d11}). The second swallow tail representing phase coexistence in the metastable portion of the large black hole branch is clearly visible.}

\label{d12}
\end{figure}

\begin{figure}
\includegraphics[width=0.5\textwidth, height=6cm]{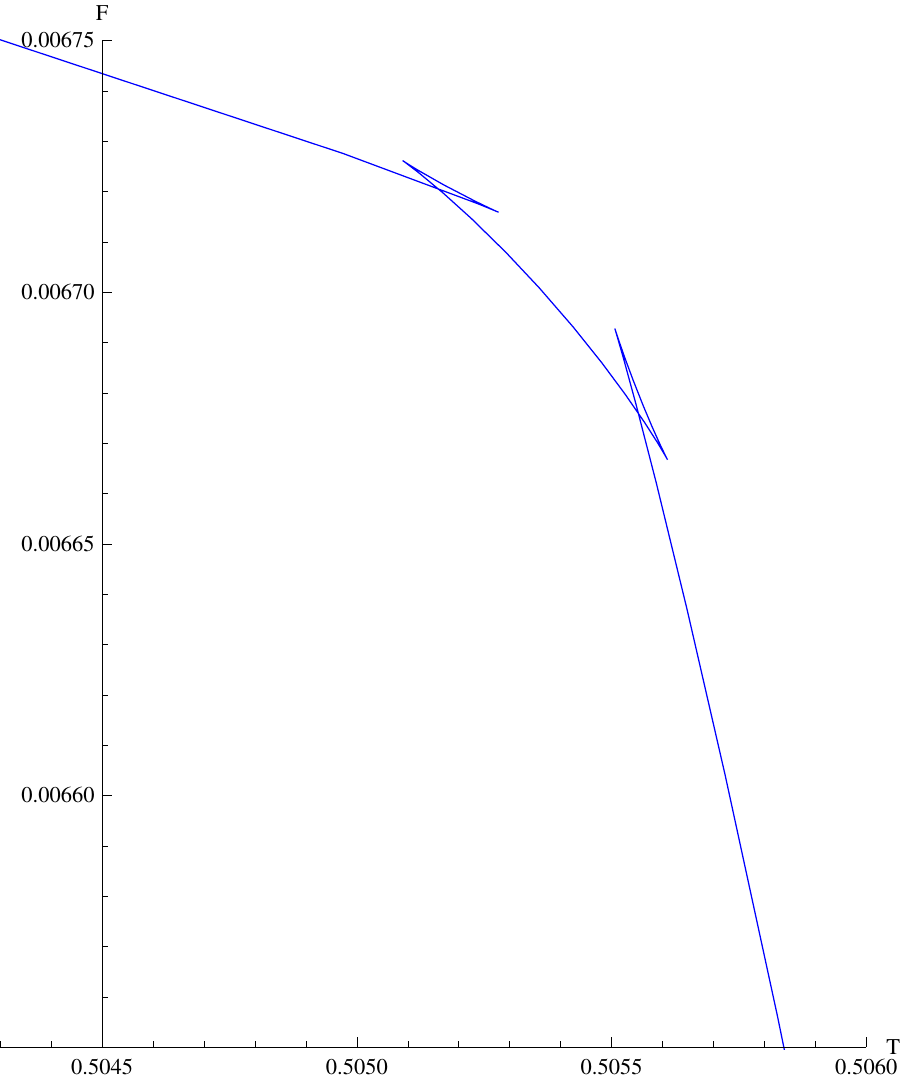}
\caption{Multicriticality in the canonical ensemble of $d=6$ GB-AdS black holes. A plot of the Helmholtz energy $F$ vs $T$ (with $\alpha=0.0822$ and $Q=0.00411$) reveals multiple phase coexistence regions transitioning from small to intermediate to large black hole.  }

\label{d22}
\end{figure}

  However, it is the Helmholtz free energy $F$ vs $T$ plots that reveal the precise picture of local stability. Thus, sometimes the sometimes the `second' swallow tail could lie on one of the locally stable small or large black hole branches (figures (\ref{d11}) and (\ref{d12})) or it might even appear as an isolated window of stability in the locally stable branch (figures (\ref{d1}) and (\ref{d2})). Apart from such cases of nested swallow tails there is also the case of separate swallow tails as in figure (\ref{d22}). Now the two phase coexistence regions follow one another. Surely when the two `humps' of a spinodal in figure (\ref{phasecan6}) attain the same height then there is a special value of $Q$ for which two critical points would exist leading multicriticality  in black hole thermodynamics. More generally, as figure (\ref{d22}) reveals, two phase transitions occur on heating the black holes for some range of parameters. Namely, there exists a small-intermediate-large black hole phase transition for these black holes. For mathematical details of phase structure and also an interesting discussion on triple points we referethe interested reader to \cite{manman}.
 
 As was discussed earlier in the case of the $d=5$ canonical ensemble, a two dimensional maneuverability in the canonical ensemble of single charged black holes essentially requires a consideration of the extended phase space scenario wherein the cosmological constant $\Lambda$ is a thermodynamic variable. However, we shall not be undertaking this interesting  analysis here since it is out of the scope of the present work. We end this section by commenting that thermodynamic geometry cannot provide insights into this very interesting phase structure since the space of independent fluctuations remains one dimensional in this ensemble.

\subsubsection{Grand Canonical Ensemble}

The grand canonical ensemble to which we now turn has a very different phase structure compared to its $d=5$ counterpart. Once again, it will be convenient to rewrite the thermodynamic functions here in terms of $\Phi$ by inverting eq. (\ref{fi6d}). The Gibbs free energy can be written as follows.
\begin{widetext}
\begin{equation}
G=-\frac{2 \pi ^2 \left(r^7+6 r^3 \alpha -144 r \alpha ^2-r^5 (1-72 \alpha )\right)+27 r^3 \left(r^2-24 \alpha \right) \Phi ^2}{12 \pi  \left(r^2+12 \alpha \right)}
\label{g6d}
\end{equation}
\end{widetext}
Similarly, the temperature becomes:

\begin{equation}
T=\frac{2 \pi ^2 \left(5 r^4+ 3 r^2+6 \alpha \right)-81 r^2 \Phi ^2}{8 \pi ^3 r \left(r^2 + 12  \alpha \right)}
\label{tphi5}
\end{equation}

The temperature vanishes at the zeroes of the following extremal function.
\begin{equation}
\mathcal{Z}_6=2 \pi ^2 \left(5 r^4+3 r^2+6 \alpha \right)-81 r^2 \Phi ^2
\label{z6}
\end{equation} 

The independent response functions relevant to the grand canonical ensemble are again obtained as three second order partial derivatives of $G$ with respect to $T$ and $\Phi$, namely, $C_\Phi$, isothermal capacitance and the thermal expansivity of the charge at constant potential,
\begin{widetext}
\begin{align}
T\left(\frac{\partial S}{\partial T}\right)_{\Phi}&=\frac{8 \pi ^2 r^2 \left(r^2+12 \alpha \right)^2 \left(2 \pi ^2 \left(5 r^4+3 r^2+6 \alpha \right)-81 r^2 \Phi ^2\right)}{3 \left(2 \pi ^2 \left(5 r^6 -3 r^4 (1-60 \alpha ) +18 r^2 \alpha  -72 \alpha ^2\right)+81 r^2 \left(r^2-12 \alpha \right) \Phi ^2\right)}\label{cphi6}\\\left(\frac{\partial Q}{\partial \Phi}\right)_{T}&= \frac{18 r^3 \left(2 \pi ^2 \left(5 r^6 -3r^4 (1-60 \alpha ) +18 r^2 \alpha -72 \alpha ^2 \right)+81 r^2 \left(7 r^2+60 \alpha \right) \Phi ^2\right)}{\pi  \left(2 \pi ^2 \left(5 r^6 -3 r^4 (1-60 \alpha ) +18 r^2 \alpha -72 \alpha ^2 \right)+81 r^2 \left(r^2-12 \alpha \right) \Phi ^2\right)}
\label{cap6}\\
\left(\frac{\partial Q}{\partial T}\right)_{\Phi}&=\frac{432 \pi ^2 r^4 \left(r^2+12 \alpha \right)^2 \Phi }{2 \pi ^2 \left(5 r^6 -3 r^4 (1-60 \alpha ) +18 r^2 \alpha -72 \alpha^2 \right)+81 r^2 \left(r^2-12 \alpha \right) \Phi ^2}\label{exp6}
\end{align}
\end{widetext}

The response functions diverge at the zeroes of the following function as is obvious from the expressions above.
\begin{align*}
\mathcal{N}_6 = 2 \pi ^2 & (5 r^6 -3 r^4 (1-60 \alpha ) +18 r^2 \alpha -72 \alpha ^2 )\\&+81 r^2 (r^2-12 \alpha ) \Phi ^2.
\label{n6}\numberthis
\end{align*}

\begin{figure}
\includegraphics[width=0.5\textwidth, height=7cm]{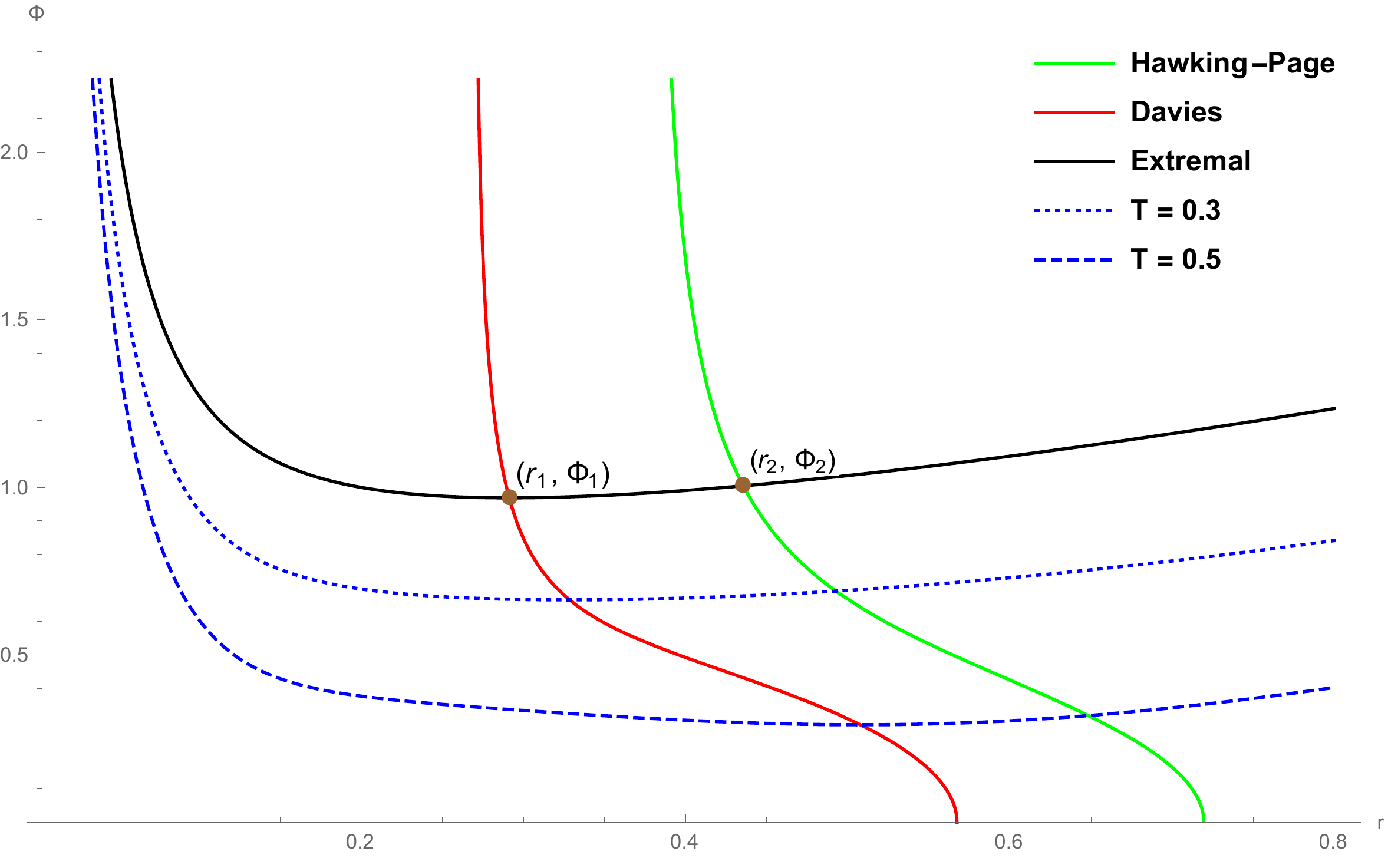}
\caption{Phase structure of GB-AdS black holes for grand canonical ensemble in $d=6$ and $\alpha=0.0060$. Red coded Davies curve determines the local stability while the green coded Hawking-Page curve fixes the global stability. Black line shows the extremal curve and both the blue coded dotted lines are isotherms at different temperatures. The values of coordinates labeled are: $(r_{1}, \Phi_{1})=(0.2913, 0.9684)$ and $(r_{2}, \Phi_{2})=(0.4356, 1.0042)$}
\label{phasegrand6}
\end{figure}

In figure (\ref{phasegrand6}), we plot in the $\Phi-r$ plane the phase structure in the grand ensemble for a fixed value of $\alpha$. The red coded curve is the locus of Davies transition points so that on crossing it from left to right the heat capacity $C_\Phi$ changes sign from negative to positive through an infinite discintinuity. The green coded curve is the Hawking-Page curve (zeros of $G$) so that the black hole is globally stable to the right of it. The black curve is the extremal curve ($T$=0), so that $T<0$ region lies above it. Also shown are two isotherms for reference. We see that there is no phase coexistence behaviour leading to criticality like its grand canonical counterpart in $d=5$. Rather, the only instability is the Davies transition between the locally unstable small black hole and the stable large black hole. Nevertheless, the phase structure, as we shall summarize in the following,  is noticeably different from the grand canonical ensemble of the RN-AdS case \cite{cliff,sahay2,sahay3}. 
 
  One can observe different thermodynamic behaviours for three ranges of $\Phi$ which we shall presently discuss. From the equations (\ref{g6d}), (\ref{n6}) and (\ref{z6}) above we obtain the following conditions. The minimum of the extremal curve in the figure above intersects the Davies curve at $r_1=\left(\frac{6}{5}\right)^{1/4} \alpha ^{1/4}$. Similarly, the extremal curve and the Hawking-Page curve intersect at $ r_2=6^{1/4} \alpha ^{1/4}$. These translate into the following values for $\Phi$.
\begin{align*}
\Phi_1 &=\frac{2^{1/4} \pi  \sqrt{\left(\sqrt{2}(1+2 \sqrt{30 \alpha })\right) } \sqrt{\left( \sqrt{30} +20 \sqrt{\alpha} \right)}}{3\ \sqrt{ \left( \sqrt{90} \left( 1 +2 \sqrt{30 \alpha} \right) \right)}}\\
\Phi_2 &=\frac{2^{1/4} \pi  \sqrt{\left(\sqrt{2}(1+2 \sqrt{6 \alpha })\right) }}{3\ \sqrt{3}} \numberthis\label{p12}
\end{align*}

\begin{figure}
\includegraphics[width=0.5\textwidth, height=6cm]{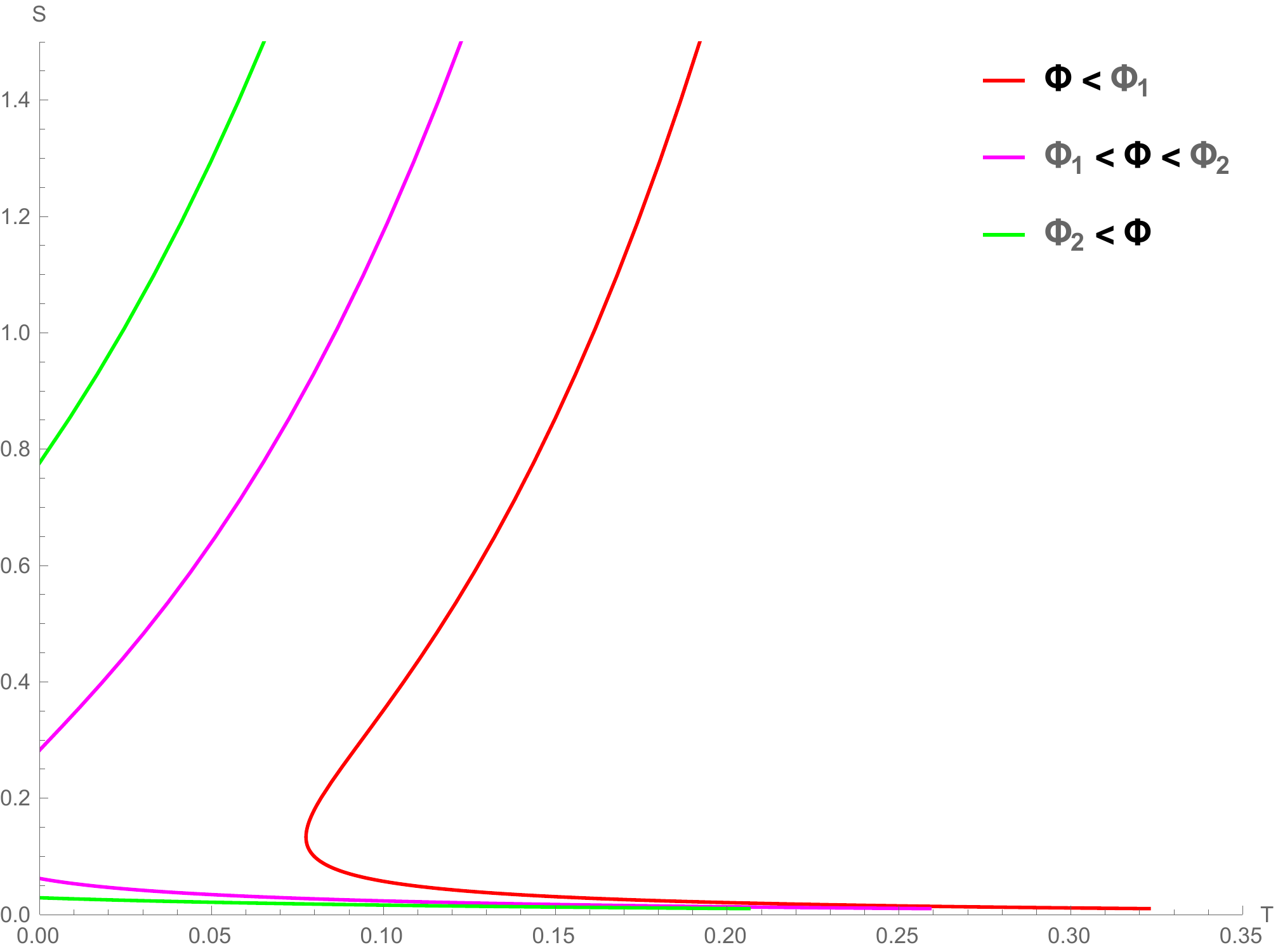}
\caption{Equipotential curves in the $S-T$ plane of the GB-AdS black holes in the grand canonical ensemble for $d=6$ and $\alpha =0.0060$. The values of potentials are: $0.9000$ (red), $0.9850$ (magenta) and $1.0500$ (green).}
\label{svstin6}
\end{figure}

Thus, for $\Phi<\Phi_1$ the the black hole exhibits a regular Davies point between the unstable and the stable branches at a finite temperature. The metastable large black hole stable branch becomes globally stable after the Hawking Page transition. For $\Phi>\Phi_1$ the stable and the unstable branches develop a `mass' gap and the Davies point disappears. Furthermore, both branches start at extremality. For $\Phi\geq \Phi_2$ the LBH branch is always preferred over the thermal AdS. In figure (\ref{svstin6}) we show representative isopotential plots in the $S-T$ plane while in figure (\ref{gvstin6}) we represent the same plots in the $G-T$ plane. Note from the last figure that the locally unstable small black hole is also globally unstable.

\begin{figure}
\includegraphics[width=0.5\textwidth, height=6cm]{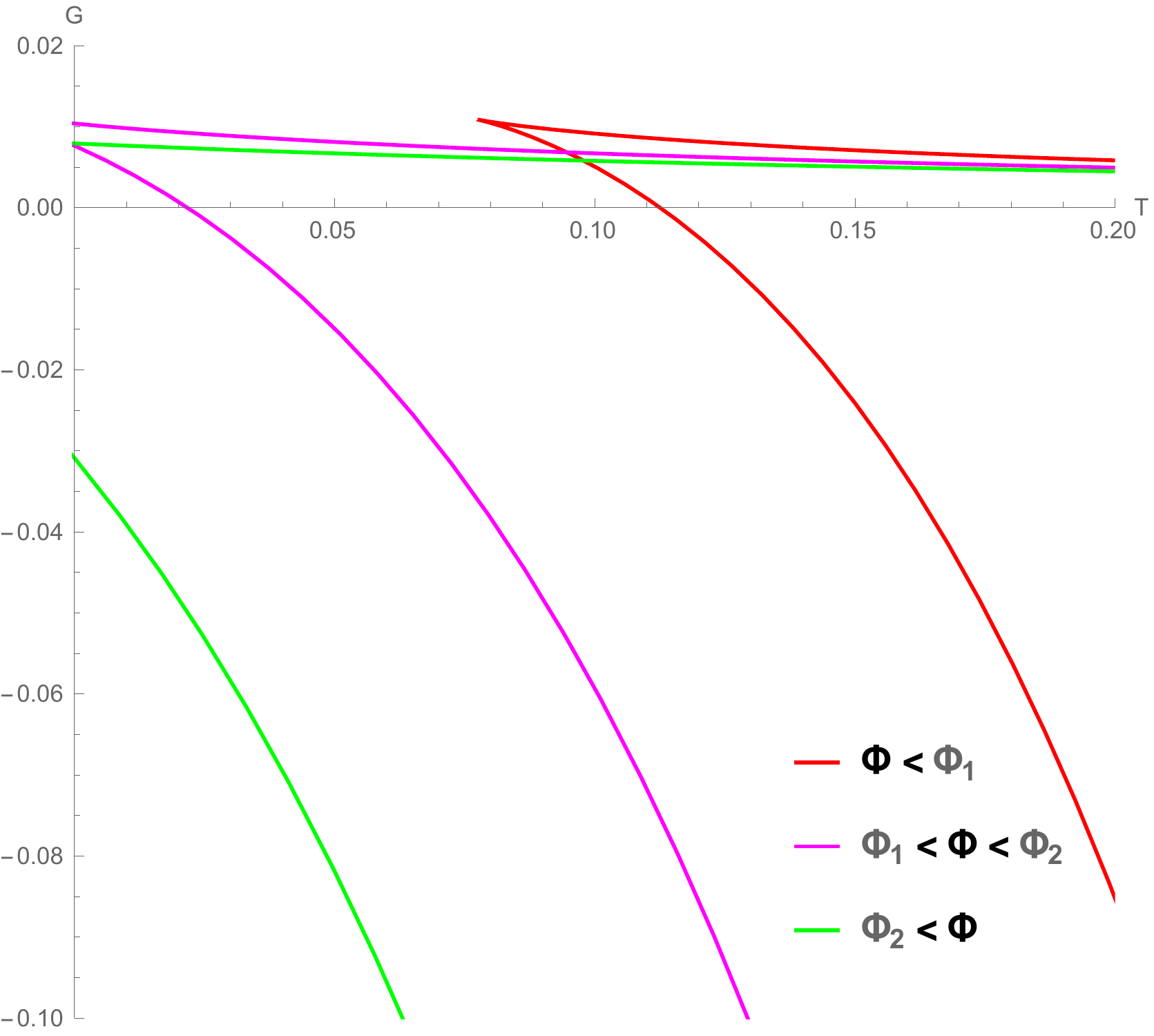}
\caption{Gibbs free energy vs temperature curve for the $d=6$ GB-AdS black holes in grand canonical ensemble with $\alpha = 0.0060$. The values of potentials are: $0.9000$ (red), $0.9850$ (magenta) and $1.0500$ (green).}
\label{gvstin6}
\end{figure}

\begin{figure}
\includegraphics[width=0.5\textwidth, height=6.5cm]{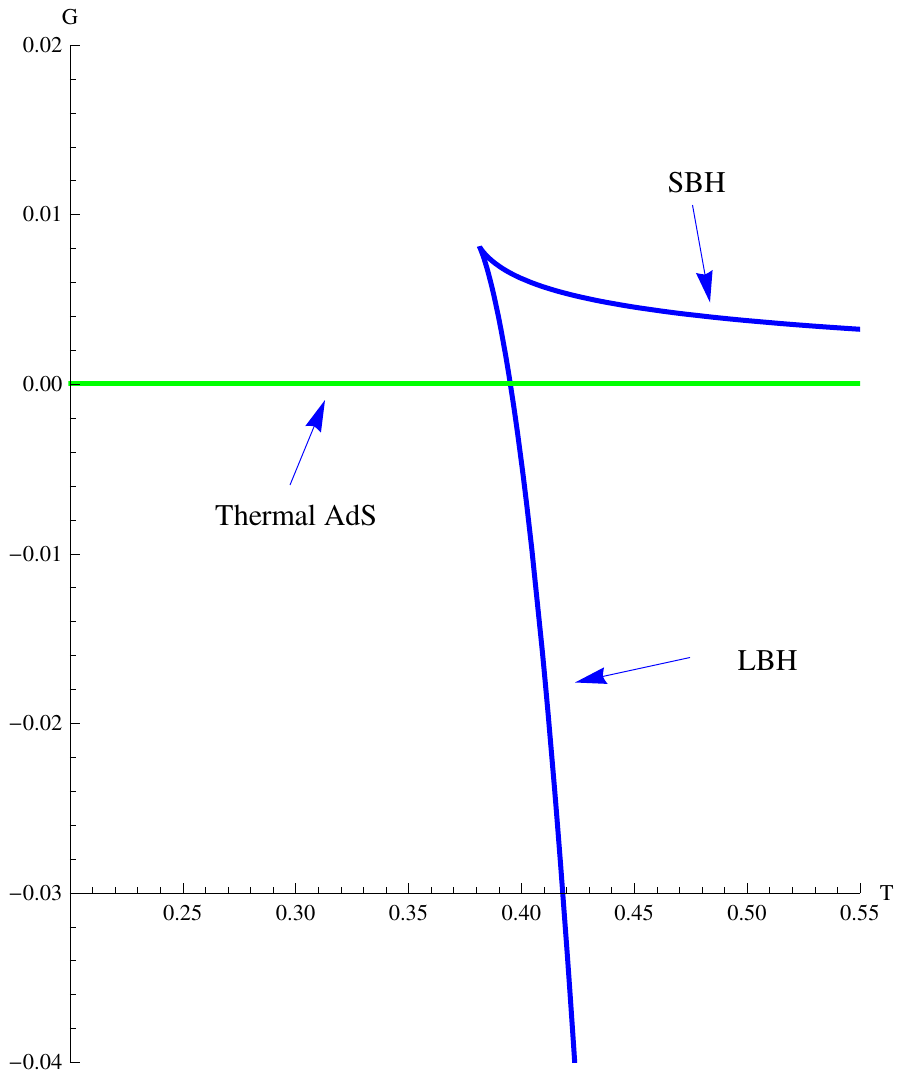}
\caption{A plot of Gibbs free energy vs. temperature representing the Hawking-Page transition as a first order transition. The line $G=0$ represents the thermal AdS and the free energy curve of the black hole is for $\Phi=0.5500$ and $\alpha=0.0060$. The Davies point is the spinodal point on the left edge of the swallow tail. }
\label{gvstin6}
\end{figure}
   
 Though it is not uncommon to refer to the Davies point as the Davies `critical' point, evidently it is a somewhat unusual critical point on the following account. While for the isolated critical points the phase transition proceeds via a continuous transformation of one stable phase into another, in the Davies case, at least when the ensemble is not microcanonical, the smaller black hole branch is thermodynamically unstable. Moreover, at temperatures below the Davies point, no black hole solutions exist.  This, therefore precludes the possibility of a `phase coexistence' region, and, at the least, the critical exponent $\beta$ becomes irrelevant. The qualitative difference with the isolated critical point therefore reflects in the critical exponents near the Davies critical point. In the usual manner, the scaling of all thermodynamic functions can be ascertained via the scaling of the singular part of the Gibbs free energy along the isotherm and the isopotential passing through the Davies singularity. The scaling of the singular part of $G$ is given as
\begin{align*}
 G_s &\sim (\Phi-\Phi_D)^{\frac{3}{2}}\,\,\mbox{along isotherm, and}\\
 G_s &\sim (T-T_D)^{\frac{3}{2}}\,\,\mbox{along isopotential.} \label{gscaledavies}\numberthis
\end{align*}
 and is expected since at the Davies point  $\partial T/\partial r=0$ but $\partial^2 T/\partial r^2$ is not (see \citep{majhi}).

 Being a turning point in the $T-r$ plane with $\partial T/\partial r =0$ and not an inflection point as the isolated critical point is, the Davies singular point is therefore in the same class as the spinodal point where the relevant response functions change sign through an infinite discontinuity. In fact, it is easy to realize this observation by a consideration of the black hole system plus the thermal AdS  as one composite system. Revisiting isopotential $G$ vs.$T$ plots in figure (\ref{gvstin6})  we set the free energy of thermal AdS (which, being the reference background, has $G=0$) against that of the black hole and observe a typical "swallow-tail" feature which is a signature of phase coexistence. While the right edge of the tail joins at $T=\infty$ and the left edge is the Davies point, the Maxwell construction point is the site of Hawking-Page transition. For the composite black hole plus thermal AdS system, while the equation of state of the black hole is known explicitly, that of the thermal AdS can be written down depending on the mix of particles we want to put into it \citep{page}.  It turns out, even without calculating the equation of state for the thermal AdS, thermodynamic geometry will have something to speak about it via the anticipated equality of $R$ at Hawking-page phase transition. We shall discuss this in the next section on geometry.
 
We stress that we do not mean to suggest that the Davies point is not a genuine transition point even if it lies in the globaly unstable region. In fact, so does the isolated critical point, a main focus of this paper. Metastable states are a matter of fact in ordinary systems. In the AdS/CFT context,  the Davies point of large black holes corresponds to the Bose condensation point in the dual field theory \cite{gubser}.

Having reviewed and supplied our perspective on the phase structure of the $\alpha'-$corrected black holes, we now proceed to set up their information geometry.

\section{Ruppeiner Geometry of Gauss-Bonnet adS Black Holes}

The starting point for a macroscopic theory of thermodynamic fluctuations is Einstein's fluctuation theory where the probability $P$ of a macroscopic state depends on its entropy as 
\begin{equation}
P= C\, \mbox{exp}\left[\frac{S}{k_B}\right].\label{einstein}
\end{equation}

Though just an inversion of Boltzmann's formula, the logic of Einstein's fluctuation theory is diametrically opposite since in the above equation information flows leftwards from a macroscopically measurable and calculable thermodynamic quantity to a statistical measure of the multiplicity of microstates. Ruppeiner \cite{rupp0} was able to covariantize the information content in the equation above by a consideration of the negative Hessian of the entropy as the Riemannian metric permeating the thermodynamic state space,
\begin{equation}
g_{\mu\nu}=-\frac{\partial^2 S}{\partial X^\mu \partial X^\nu}.
\end{equation}

 It turns out, such a reorganization of information into invariants of geometry like the geodesics or the curvature tensor offers several new insights. In particular, the scalar curvature directly connects to the underlying correlations in the microscopic degrees of freedom. The connect is much more pronounced in systems, or in certain regimes of systems like the critical region, where the statistical correlations are strong. It has been successfully verified for several cases that in such regimes the absolute value of the inverse curvature, namely the curvature length squared of the thermodynamic manifold, is equal to the dynamical (or the singular) part of free entropy density of the system modulo an order unity constant. On the other hand, the sign of $R$ is related to the nature of underlying ineteractions. In the convention prevalent in TG literature\footnote{Namely, Weinberg's convention \cite{wein} wherein $R$ for the two-sphere is negative.}, $R$ is negative when long range attractive interactions dominate, like near the critical point, in bosons, etc, while it is positive when short range repulsive interaction dominate, like for fermions, or solid like repulsive interactions for fluids at high density or even in some regions at low enough densities like the dense vapor phase \cite{rupp3,rupp5}. In the vicinity of the critical point, where correlations are still strong and hence $R$ large enough, the asymptotic coexistence curve can be obtained by setting the curvature for both phases to be equal. This has been termed the $R$-crossing method of obtaining the coexistence region and complements the Maxwell construction \cite{rupp2, rupp4}. In fact, as shown in \cite{rupp4} the $R$-crossing method of obtaining phase coexistence goes beyond the mean-field Maxwell constructions and shows a very good match with coexistence curves obtained by simulation of Lennard-Jones fluids. This is a further confirmation of the correspondence of $R$ with the correlation volume for the extensive systems since the coexisting phases are expected to have the same length scale of density fluctuations. Furthermore, $R$ stands out as a unique marker of the Widom line in the supercritical region \citep{rupp2, rupp4}.
 
 In the following we shall calculate in order the scalar curvature for the $d=5$ and the $d=6$ Gauss-Bonnet-AdS black holes and attempt to verify the brief description of $R$ outlined above in the context of black holes.
 
\subsection{d=5 case} 

The calculation of $R$ proceeds in a standard way via a calculation of the Hessian metric relevant to the $5 -d$ black holes. To re-emphasize, the geometry thus obtained represents the black hole in a grand canonical ensemble. This is simply because in the Taylor expansion of the entropy about its maximum value at equilibrium no constraint has been imposed on the variation of the extensive variables $M$ and $Q$ \cite{sahay}. The two dimensional thermodynamic metric contains the same information as the three independent response functions  outlined in a previous section in eq. (\ref{cphi5}) $-$ eq. (\ref{exp5}). The scalar curvature for the $d=5$ case can be obtained from the metric and is expressed as

\begin{align}
R_5=\frac{\mathcal{P}_5}{3 \pi ^2 \left(r^2+4 \alpha \right)^2\, \mathcal{Z}_5\,\mathcal{N}_5^{\,2}}\label{R5}
\end{align}

where the numerator $\mathcal{P}_5$ is

\begin{align*}
&\mathcal{P}_5&\\ &=4 r \,[27 \pi ^6 (12 r^{10}+48 (1-4 \alpha ) \alpha ^2-16 r^8 (1-23 \alpha )\\&+4 r^2 \alpha  (1+36 \alpha +128 \alpha ^2)-12 r^4 \alpha  (1-48 \alpha -320 \alpha ^2)\\&-3 r^6 (1+16 \alpha -832 \alpha ^2))\\&+3456 \pi ^4 (4 r^8-24 \alpha ^2 (1-8 \alpha )-r^6 (1-56 \alpha )\\&-4 r^4 \alpha  (1-56 \alpha )-2 r^2 \alpha  (1+8 \alpha -256 \alpha ^2)) \Phi ^2 \\&+ 36864 \pi ^2 (3 r^6+12 r^4 \alpha +4 r^2 (1-4 \alpha ) \alpha \\&+16 (3-20 \alpha ) \alpha ^2 ) \Phi ^4-1048576 \alpha  (r^2+12 \alpha ) \Phi ^6]\numberthis
\label{P5}
\end{align*}
and the functions $ \mathcal{Z}_5$ and $\mathcal{N}_5$ where defined earlier for the $d=5$ case. Clearly, from the presence of $\mathcal{N}_5$ in the denominator it can be explicitly seen that $R$ diverges along the spinodal corresponding to the grand canonical ensemble. In particular, $R$ is therefore also singular at the critical point in the grand ensemble. Similarly, from $\mathcal{Z}_5$ in the denominator it can be checked that $R$ diverges to positive infinity at the extremal point if it exists for the black hole. Also, from the factor of the horizon radius $r$ in the numerator in eq. (\ref{P5}) above it can be checked that $R$ goes to zero from the positive side whenever the black hole horizon shrinks to zero.

Now we undertake a detailed observation and analysis of features of thermodynamic geometry for $d=5$ as narrated by the curvature of the thermodynamic manifold. Keeping in mind figure (\ref{phasegrand5}) and the discussion around eq. (\ref{rcr5}), we divide the phase-space in the $\Phi-r_h$ plane into three parts. Part $A$ has $0<\Phi<\Phi_{cr}$. Thus this part comprises the coexistence region culminating in the critical point. Part $B$ has $\Phi_{cr}<\Phi<\Phi_0$ which is the part of the beyond-critical region where there is no extremality yet. Part $C$ has $\Phi>\Phi_0$ onward where extremal black holes exist. 

\begin{figure}
\includegraphics[width=0.5\textwidth, height=7cm]{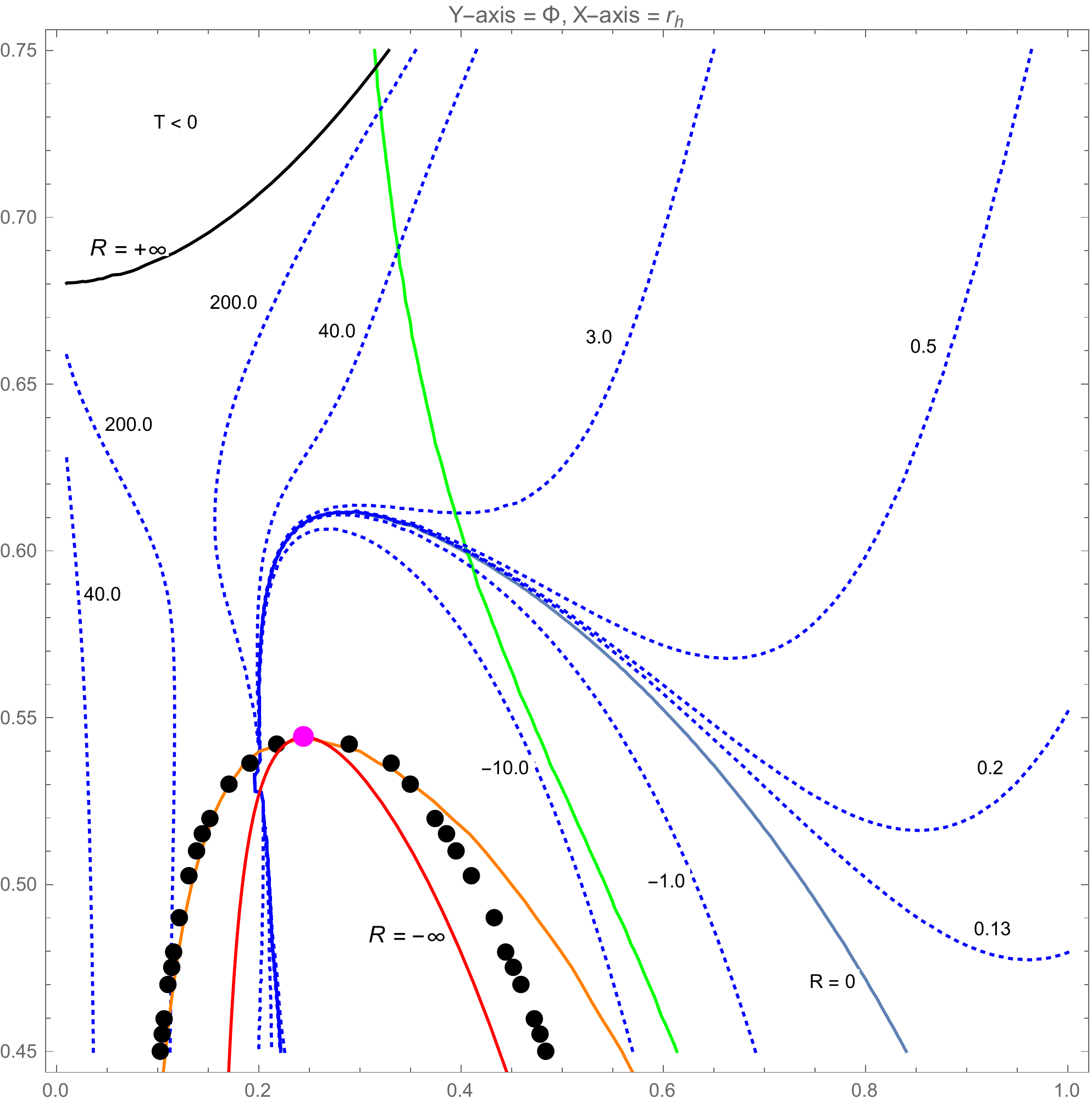}
\caption{Plot in the $r_h-\Phi$ plane of the grand canonical ensemble phase structure and the contours of $R$ for the GB-AdS black holes in $d=5$ for $\alpha=0.0050$. The red curve is the spinodal line, the black one is the extremal curve, the blue one is the locus of vanishing $R$ and the green is the Hawking-Page line. The phase coexistence curve obtained via Maxwell construction is orange coded. The magenta dot is the critical point. The black dots represent phase coexistence via $R$-crossing. The dotted blue curves are contours of $R$ with the values indicated.}
\label{Rsign}
\end{figure}

 In figure (\ref{Rsign}) we plot in the $r-\Phi$ plane the features of $R$ as superimposed over the phase structure in the grand canonical ensemble for $\alpha=0.005$ for  the critical potential works out to, $\Phi_{cr}=0.544$. The black colored curve in the top left is the extremal curve (``T-curve") above which the temperatuer is negative, the blue curve is the zero of $R$ ("R-curve"), the red curve is the spinodal curve ("S-curve") along which $R$ becomes infinite, the green curve is the phase coexistence curve ("M-curve") obtained va the Maxwell construcion while the back dots are the coexistence curve as obtained by the $R$-crossing method ("R-max"). Also plotted are several blue dotted curves representing contours of $R$. Let us focus on part A or the subcritical region. A characteristic feature of $R$ here is that it remains positive for much of the stable small black hole branch. For the large black hole branch its behaviour is similar to the  $4$-dimensional RN-AdS case \cite{sahay2} where it remains large and negative in regions close to the spinodal and then turns to small positive values (less than unity) before decaying to zero away from the phase coexistence region. Interestingly there is a close match of the R-curve with the M-curve in the small black hole phase but not as good in the large black hole phase. We note that since there is a sign difference of $R$ along much of the two branches, we have equated the absolute values of $R$. The sign difference between the small and the large branches could possibly indicate a fundamental difference in the underlying statistical correlations of one black hole branch as compared to the other. Through a simple minded juxtaposition of the results in \cite{rupp3} for fluids we might tentatively propose that the small black hole branch is solid like while the large black hole branch is fluid like. Of course, since nothing concrete is known of the microscopic details, this statement is just a mnemonic for the anticipated difference in the underlying statistical features of the two branches. We shall comment on the significance of state space geometry for the Hawking Page in a subsequent passage below.

Crucially, the fact that the sign of $R$ becomes negative in the neighborhood of the the critical point shows consistency with Ruppeiner's scaling arguments that in the critical region $R$ always remains negative \citep{rupp}. Thus, in particular, $R$ diverges to negative infinity at the critical point. In keeping with the scaling approach we choose to call the vicinity of the critical  point where $R$ remains negative and sufficiently large as the `asymptotically critical region'. Since a negative $R$ is consistent with its scaling form valid in the near critical region we may reasonably suggest that the crossing of $R=0$ curve with the phase coexistence curve sets the boundary of asymptotically critical region. However, we shall qualify this assertion soon when we briefly compare with the geometry of van der Waals fluid.

We now check for the scaling behavior of $R$ near the critical point in the supercritical region. The analysis remains exactly analogous to the expansion of the Gibbs free enery near the critical point in eq. (\ref{gstherm}) and eq. (\ref{gspot}). We expand $1/R$ in the supercritical region along the critical isopotential and the critical isotherm respectively to get the following.
 
 \begin{align*}
 R_5^{-1} &= -\frac{1152 \sqrt{3}\ 2^{1/3} \pi ^{10/3}\ \alpha ^{11/6} }{1-72 \alpha }\,\left(T -T_{cr}\right)^{4/3}+ \ldots\\
 R_5^{-1} &= -\frac{1152 \sqrt{3}\ 2^{1/3}  \pi ^{2/3}\ \alpha^{7/6} \left(1 -72 \alpha\right)^{2/3} }{1-72 \alpha }\,\left(\Phi -\Phi_{cr}\right)^{4/3}+\ldots\label{Rscaling}\numberthis
 \end{align*}
 
 Indeed, the scaling behaviour of $R$ is the same as that of the singular part of the Gibbs free energy. Significantly, the ratio of the amplitudes of $R$ and the corresponding ones of $G_s$ also match up to the same (coupling-constant-dependent-value) $\frac{32 \sqrt{3} \pi  \sqrt{\alpha }}{1-72 \alpha }$, thus confirming the equality of $R$ with the inverse of $G_s$ modulo a constant. Further, following Ruppeiner's analysis \citep{rupp}, we now check for the consistency of $R$ with two-scale factor universaliy by observing its scaling along the critical iscoharge (which is analogous to the isochore in the fluid case) in the supercritical region. 
 
 We recall that the scaling form of $R$ in the critical region (from the supercritical side) is given as follows \cite{rupp}.
 
\begin{equation}
R=\frac{(\beta\delta-1)(\delta-1)\,T_{cr}}{(\beta\delta-1+\beta)(\delta+1)Y(0)}\,t^{\alpha-2}
\label{scale1}
\end{equation} 
where $\alpha,\beta$ etc are critical exponents, $t$ is the reduced temperature and $Y(z)$ is the function of one variable appearing in the scaling expression of the singular free energy. $Y(0)$ is its zero  field value. Similarly, the heat capacity at constant density (charge in this case) has the following scaling form
\begin{equation}
C_\rho=-\frac{(2-\alpha)(1-\alpha)Y(0)t^{-\alpha}}{T_{cr}}
\label{scale2}
\end{equation}

The above equation constrains $Y(0)$ to be negative since $C_\rho$ is positive at the critical point.

The interpretation of $R$ as the correlation volume in ordinary systems is further confirmed  by its connection two-scale factor universality \cite{rupp}. According to two-scale factor universality, given that the correlation length is $\xi=\xi_+t^{-\nu}$, where the $`+'$ indicates supercritical region, and the heat capacity goes as $C=C_+t^{-\alpha}$, the combination $\xi^d C t^2$ is a universal constant \cite{stauff}. The $t-$independence of the combination follows from the hyperscaling exponent relation $\nu d = 2-\alpha$. Following eq. (\ref{scale1}) and eq. (\ref{scale2}) above, the consistency of $R$ with two scale factor universality is easily established as follows.
\begin{equation}
R C_\rho t^2=-\beta(\delta-1)(\beta\delta-1).
\label{twoscale}	
\end{equation}

For mean field model like the van der Waals gas, the above relation works out as follows \citep{rupp}.
\begin{equation}
R C_\rho t^2=-\frac{1}{2}.
\label{twoscalevdw}
\end{equation}

Thus, $R$ not only scales as the correlation volume  near the critical point but its amplitude too is the same. A brief comment is in order regarding eq. (\ref{twoscalevdw}) above to further recommend $R$ as correlation volume. The first order correction to the mean field theory to account for local inhomogeneity is realised by adding a square gradient term to the free energy. This results in a well known correlation length expression which scales as
$\xi\sim t^{-1/2}$ along the critical isochore in the supercritical region. As can be checked, this expression for correlation length is {\it not} consistent with two-scale factor universality. This does show that the calculation of $R$ even within a mean field context supplies a correction to the theory which does better than the standard square gradient correction in the near critical region.

 Now, coming back to black holes for which no direct notion of correlation length exists, it will be interesting to implement the above relations. After some algebra, the expansion of $R$ along $Q=Q_{cr}$ works out to
 \begin{align}
 R^{-1}=\frac{64 \pi ^4 \alpha ^{3/2} \left(T-T_{cr}\right)^2}{\sqrt{3} (-1+72 \alpha )}+\ldots
 \end{align}
while the expansion of $C_Q$ works out to
 
 \begin{align}
 C_Q=\frac{1152 \sqrt{3} \pi ^2 \alpha ^{5/2}}{1-72 \alpha }+\frac{192 \pi ^3 \alpha ^2 (1+108 \alpha ) \left(T-T_{cr}\right)}{(1-72 \alpha )^2}+\ldots
 \end{align}
 
 It can be easily verified from the above equations that for the Gauss-Bonnet AdS black holes the two scale factor combination gives
 \begin{equation}
 R C_Q t^2=-\frac{1}{2}
 \end{equation}
 
 Perhaps it is not unexpected since the black hole too has a mean field equation of state in the critical region as shown in eq. (\ref{criteos}). At the same time it strongly suggests a correlation length like role for $R$ in the critical region of the black hole. We restate the assertion made at  the beginning of this paper and earlier in \cite{sahay} that the (absolute value of) thermodynamic scalar curvature is a measure of the {\it number of uncorrelated domains} in black holes. This identification gets better in the near critical region. By a domain, we mean here a bunch of statistically correlated microscopic degrees of freedom much like in a ferromagnet with a key difference being that in our assertion we do away without any mention of the relative location of these degrees of freedom. At the same time, in the spirit of the holographic principle, on could imagine these to  be located on the black hole horizon. 
 
Given the scaling equation for $C_\rho$, eq. (\ref{scale2}) the scaling function $Y(0)$ for the Gauss-Bonnet case works out to
 \begin{equation}
 Y(0)=-\frac{10368 \pi  \alpha ^3}{1-72 \alpha }
 \end{equation}

Surely, $Y(0)$ remains negative for all values of $\alpha$ that permit phase coexistence, namely $\alpha<1/72$.

 A useful comparison can be made with the sign change of $R$ in the van der Waals model{\footnote{This line was suggested by George Ruppeiner in a private conversation}. In figure (\ref{vdw}), we plot in the plane of reduced temperature $t$ and reduced molar volume $\sigma$, the red coded spinodal curve corresponding to the universal van der Waals equation of state. The critical point is at $t=\sigma=1$. The blue coded curves labeled $R1,R2$ and $R3$ are the zeroes of $R$ for the constant volume specific heat capacity  $C_1=4.5,C_2=2.5$ and $C_3=0.5$ respectively. $R$ is positive under each of these respective `zero' curves. It can be seen that for simple molecules (low heat capacity), $R$ remains negative for both phases. However, for more complex molecules (larger heat capacity), $R$ becomes positive at low temperatures and large densities, thus, signalling solid like features in the particular regime. This qualitative picture offered by the geometry of a simple mean field model has much to recommend it as is borne out by simulation studies of Lennard$-$Jones fluids \cite{rupp3}. For the Gauss$-$Bonnet case, the locus of zeroes of $R$ similarly cuts the coexistence/spinodal curves to the {\it left} of the critical point in the SBH phase so that the low temperature, low potential phase has a positive curvature. Also note another similarity of the large heat capacity branch $R1$ with the black hole case. Namely, the locus of zeroes of $R$ for both cases is prominent beyond the critical point. Thus, even in the supercritical region, where the phase change is continuous without involving any singularity, $R$ sort of keeps track of the two phases by its signature. With reference to earlier discussion with reference to figure (\ref{Rsign}) about $R$ demarcating the asymptotically critical region of GB-AdS in $d=5$, we comment here that while something similar can be ascribed to the behaviour of $R$ for complex fluids in figure (\ref{vdw}), same cannot be said of simple fluids for which $R$ remains negative all along or for much of the coexistence curve.

\begin{figure}
\includegraphics[width=0.5\textwidth, height=7cm]{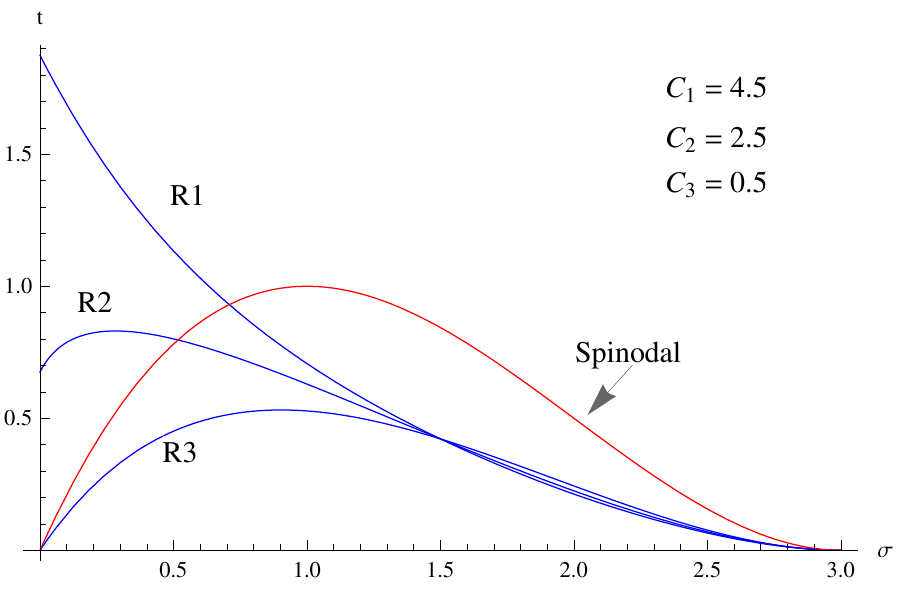}
\caption{Reduced temperature vs. reduced molar volume plot of the phase structure and geometry of the van der Waals model. The red coded spinodal is also the locus of infinities of $R$. The zeroes of $R$ for three different constant volume specific heat capacities are shown.}
\label{vdw}
\end{figure}

\begin{figure}
\includegraphics[width=0.5\textwidth, height=7cm]{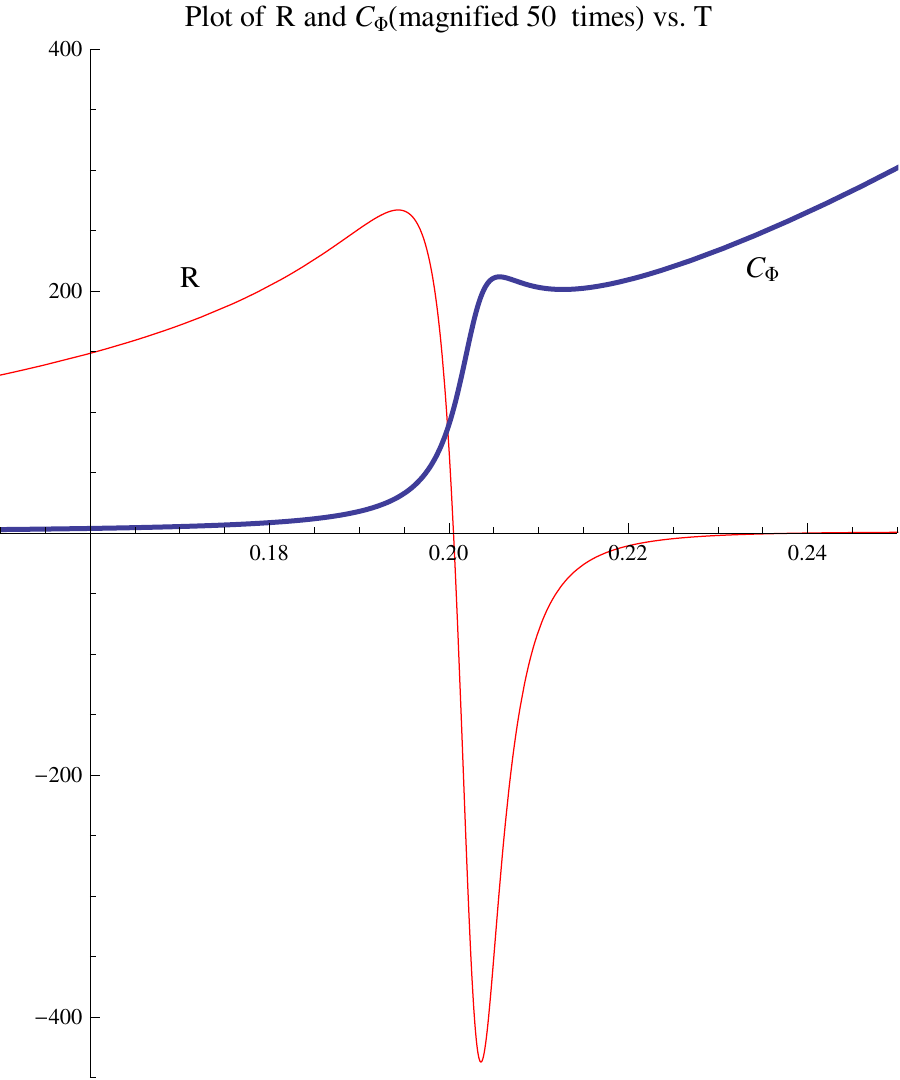}
\caption{$R$ vs. $T$ and $C_\Phi$ vs. $T$ plot in region B for $\alpha=0.0050$ and $\Phi=0.5700>\Phi_{cr}=0.5440.$ The heat capacity has been magnified 50 times for a visual comparison with the $R$ plot.  }
\label{Rsuper}
\end{figure}

\begin{figure}
\includegraphics[width=0.5\textwidth, height=7cm]{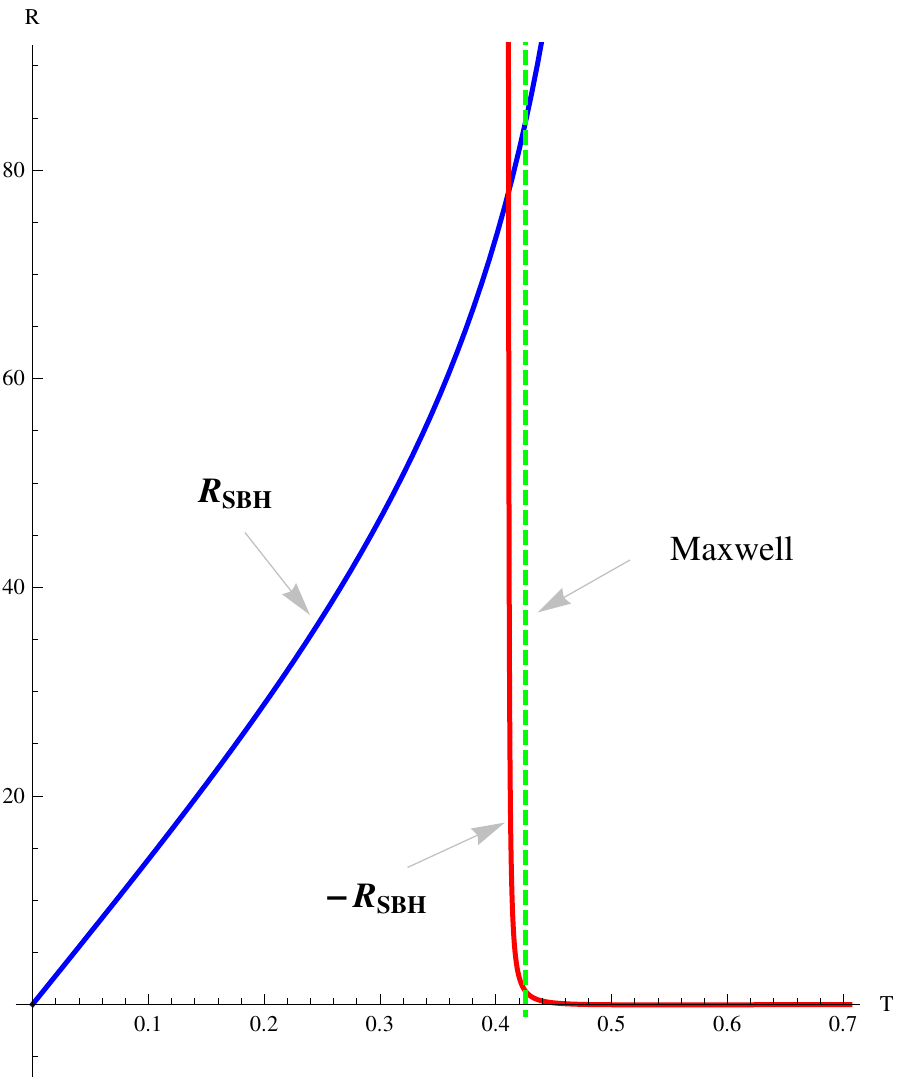}
\caption{A demonstration of the $R$-crossing method to obtain the coexistence curve in region A, with $\alpha=0.0050$ and $\Phi=0.2000<\Phi_{cr}=0.5440$}
\label{Rcross}
\end{figure}

\begin{figure}
\includegraphics[width=0.5\textwidth, height=7cm]{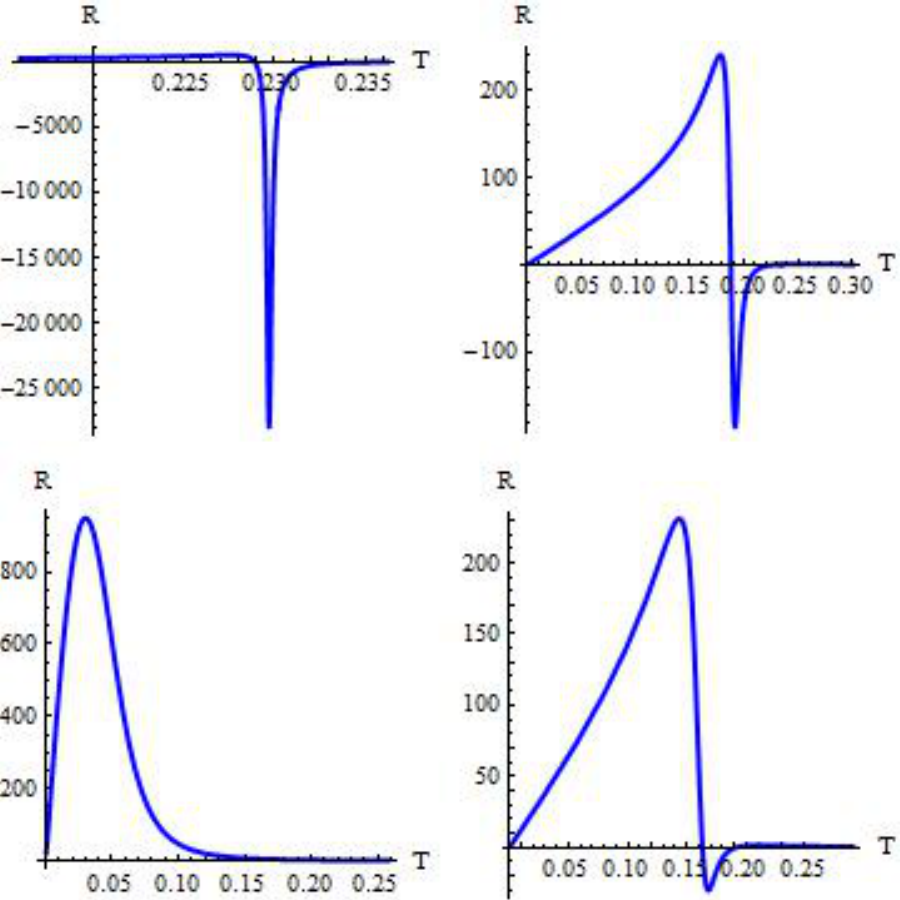}
\caption{Four plots of $R$ vs. $T$ in the region B with $\alpha=0.0050$ and $\Phi=0.5480,0.5800,0.6000,0.6600$ clockwise from top left. The region C, which contains extremal black holes, starts at $\Phi=0.6800$ while the critical value $\Phi_{cr}=0.5440$. }
\label{clockwise}
\end{figure}

 
 

\begin{figure}
\includegraphics[width=0.5\textwidth, height=7cm]{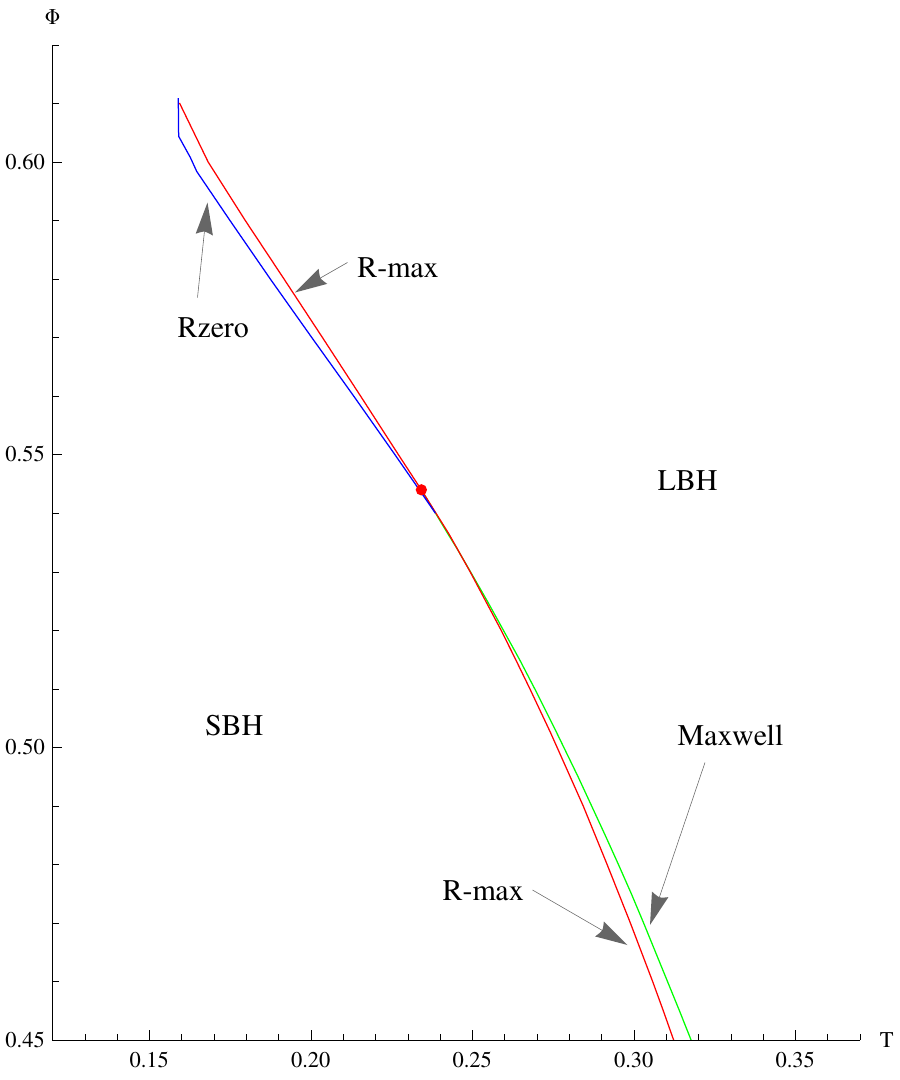}
\caption{A plot in the $\Phi-T$ plane of the phase coexistence curve as obtained by the Maxwell rule and as predicted by the $R$-crossing method. The critical point is a red dot. In the supercritical region we plot the line of zeroes of $R$ and the locus of negative troughs of $R$ as discussed in the text. Note carefully that the curve of zeroes of $R$ starts from {\it below} the critical point, which we take to be a measure of the asymptotically critical region. We have set $\alpha=0.0050$.}
\label{Rcros}
\end{figure}

We now briefly discuss regions B and C, namely the regime beyond the critical point in figure (\ref{Rsign}). Just as for the van der Waals model, the scalar curvature $R$ continues to go through negative minima of large magnitude in the supercritical region in the vicinity of the critical point. In earlier works it was shown in the context of pure fluids \cite{rupp2,rupp4} that large values of $R$ are generic in the supercritical region and unequivocally represent the Widom line, the line of correlation function maxima in the supercritical region. In the context of liquid-gas transitions in fluids, the Widom line marks a crossover from the liquid like state to the vapor-like state in the supercritical region. Not surprisingly, for the present case of black holes too, the line of extrema of $R$ continues beyond the critical point into the region B. Interestingly, the (negative) minimum of $R$ is always closely preceded by its sign change from positive to negative. This is somewhat similar to, but sharper than, the $R1$ curve in the vdW case in figure (\ref{vdw}). We assert that the sign change in tandem with the minima of $R$ roughly identifies a spread out region of crossover from the SBH$-$like phase to the LBH$-$like phase in the supercritical region.  As one moves away from the critical point towards higher $\Phi$ values in figure (\ref{Rsign}), the negative minimum of $R$ becomes more and more shallow until this feature disappears for potential values higher than the maximum of the blue coded $R=0$ curve. In this region $R$ remain positive throughout. In figure (\ref{Rsuper}) we plot on the same graph $R$ vs. $T$ as well as $C_\Phi$ (magnified 50 times) vs. $T$. We see that the sign change of $R$ closely precedes its negative dip. The sharp change in $C_\Phi$ roughly corresponds to this region. Also note the large positive extremum of $R$ $-$ a typical feature in region B. Admittedly, response coefficients (like $C_\Phi$ in the figure), which typically represent second moments of fluctuations, do not betray the positive crest of $R$. This leads us to tentatively think of this feature of $R$ in region B as a higher moment effect. Further probing this feature, we find that on increasing $\Phi$ towards region C, the positive crest of R becomes higher and moves leftwards towards zero temperature, finally reaching to positive infinity at $T=0$ in the region C which has extremal black holes. Therefore, the continuity of the positive peak in region B with its positive infinite value at extremality in region C prompts us to think of this feature that it has got a higher moment memory effect present in region B. Thinking in terms of the geometry of the thermodynamic manifold, the contour lines of $R$ at the top left of figure (\ref{Rsign}) tell us more directly of a high positive curvature region in the vicinity of a line singularity in the manifold at $T=0$. In figure (\ref{clockwise}), we show clockwise from top, four representative plots of $R$ vs. $T$ for $\Phi$ values in the supercritical region B, starting from a value close to the critical point. The general trends are easily observable. Thus, the negative dip of $R$ becomes less and less shallow as $\Phi$ increases away from its critical value until it vanishes completely. Also visible is the leftward movement towards $T=0$ and the increasing height of the positive peak as $\Phi$ approaches region C.

Admittedly, we do not understand much the positive divergence of $R$ at extremality in the region C beyond the sort of circular argument that it seems to be a higher moment effect. Simple thermodynamics, with its second order response functions like the capacitance or the heat capacity, gives no hint of a singularity at extremality. In fact it is quite the opposite. Since all the response coefficients smoothly go to zero at $T=0$ as they should, hence thermal fluctuations vanish at zero temperature. There is certainly a mechanical instability at extremality against perturbations of the horizon \cite{stefano}. The picture to bear in mind is that at extremality the black hole has the maximum possible charge or angular momentum that it can hold \citep{rupp6}. Any further addition will likely tip off the balance and possibly start an instability which will grow and might eventually break the horizon into smaller pieces. Is it possible for the state space geometry to encode this mechanical instability? We think it could likely be the case. The fact that the horizon area, a mechanical quantity, and the entropy, a thermodynamic quantity, are the same thing\footnote{They are not the same in the Gauss$-$Bonnet case but nonetheless both are simple functions of the horizon radius.} for black holes at least provides a consistency to our argument. Possibly the nature of mechanical instability could also explain the positive values of $R$ near extremality   for all black holes. We hope to engage with this important issue in future. We note the trend in the extremal limit to be
\begin{equation}
R\, C_Q \rightarrow 1\,\,\,\mbox{as}\,\,\,T\rightarrow 0.
\label{extR}
\end{equation}
which is exactly the same as for the $4-d$ Kerr Newman black holes \cite{ruppblack} and for the $4-d$ Kerr Newman AdS black holes \cite{sahay3}. Note that in \cite{ruppblack} where the above trend was first noticed, a connection was made with a similar trend in the zero temperature limit of the $2-d$ ideal Fermi gas.

 We now explore the phase coexistence region (region A) and the supercritical region (region B) around the critical point as predicted by the $R$-crossing method. Let us recall that except for the scaling regime near the critical point $R$ changes sign from the SBH branch to the LBH branch. Wherever it happens we mod away the sign with a view to matching the magnitude of curvatures along the coexistence curve. The $R$-crossing method is demonstrated in figure (\ref{Rcross}) where we plot in the $R-T$ plane the $R$ corresponding to the SBH and LBH branches with differnt color codes. The crossing of the two curves with each other naturally represents the highest curvature point for each of the two branches. For the sake of comparison, we have also indicated the temperature obtaind via the Maxwell rule. Our results are given in figure (\ref{Rcros}) where we plot in the $\Phi-T$ plane the `vapor-pressure' curve as given by the Maxwell rule in conjunction with the one obtained by the $R$-crossing method. The two curves agree well with each other close to the critical point beyond which they start moving apart.
 
  Which is a better measure of first order transitions? It is not possible to address this question yet since there is no concrete knowledge available of black hole microstates. However, to recommend the $R$-crossing method we would add that for extensive systems this method takes into account the local inhomogenieties associated with increased fluctuations near the phase coexistence region which the mean field based Maxwell construction method does not. As shown by \cite{rupp4} in the context of simple fluids, the match of $R$-crossing method with simulation generated phase coexistence curve is "very good". Hopefully a quasinormal mode analysis or an AdS/CFT check would settle the issue in future for the case of black holes. Further, in the supercritical region we see how the line of zeroes of $R$ and that of its negative minima go along together, marking a crossover region between the small and large black holes. Towards the end of this feature, the magnitude of the negative trough of $R$ is of the order unity.
  
  In light of the discussions on $R-crossing$ and the discussion around figure (\ref{gvstin6}), we now discuss the relevance of geometry to Hawking$-$Page (HP) transition. The intersection of the HP curve with the zero of $R$ in figure (\ref{Rsign}) above offers an interesting observation. For $\Phi$ values above the intersection as the Hawking Page curve moves up towards lower temperature it cuts across higher and higher positive values of $R$. Thinking of the HP transition as a first order transition between the GB thermal AdS phase and the black hole phase of the Gauss$-$Bonnet-AdS {\it spacetime}, the value of $R$ of the black hole phase at the HP transition point should roughly correspond to the $R$ of the thermal Gauss$-$Bonnet-AdS phase, since at the first order phase transition point the scalar curvatures of the two coexisting phases are thought to be equal. In this indirect manner therefore, we obtain some useful information about the underlying prevalent microscopic interactions in thermal AdS. Therefore, large positive  values of $R$ for thermal AdS in the low temperature regime are suggestive of a dominant fermionic interaction at low temperatures. We recall that for an ideal fermi gas $R$ goes to positive infinity at low temperatures \cite{ruppblack}. Interestingly in the higher temperature regime, namely for $\Phi$ values below the intersection in figure {\ref{Rsign}), the HP curve runs almost parallel to order unity negative values of $R$. This suggests a more modest ideal gas like classical behaviour at higher temperatures. We note that for the $d=5$ RN-AdS case, the Hawking Page curve does {\it not} intersect the zero of $R$ as it does for $d=4$ \cite{sahay3}. In $d=5$ RN-AdS case, the Hawking-Page curve lies entirely {\it inside} $R=0$ curve in the negative $R$ region.

\subsection{d=6 case}

The thermodynamic geometry for the $d=6$ case is obtained in an exactly similar manner. Once again, the Ruppeiner metric has three independent components which contain the same information as the three response coefficients relevant to the grand canonical ensemble, as given in eqs. (\ref{cphi6}), (\ref{cap6}) and (\ref{exp6}). The state space curvature expressed in $\Phi-r$ co-ordinates becomes 
\begin{equation}
\mathcal{R}_6= \frac{\mathcal{P}_6}{ \pi ^2 r^2 \left(r^2+12 \alpha \right)^2\,\mathcal{Z}_6 \,\mathcal{N}_6^{\,2}}
\end{equation}
\begin{figure}
\includegraphics[width=0.5\textwidth, height=6cm]{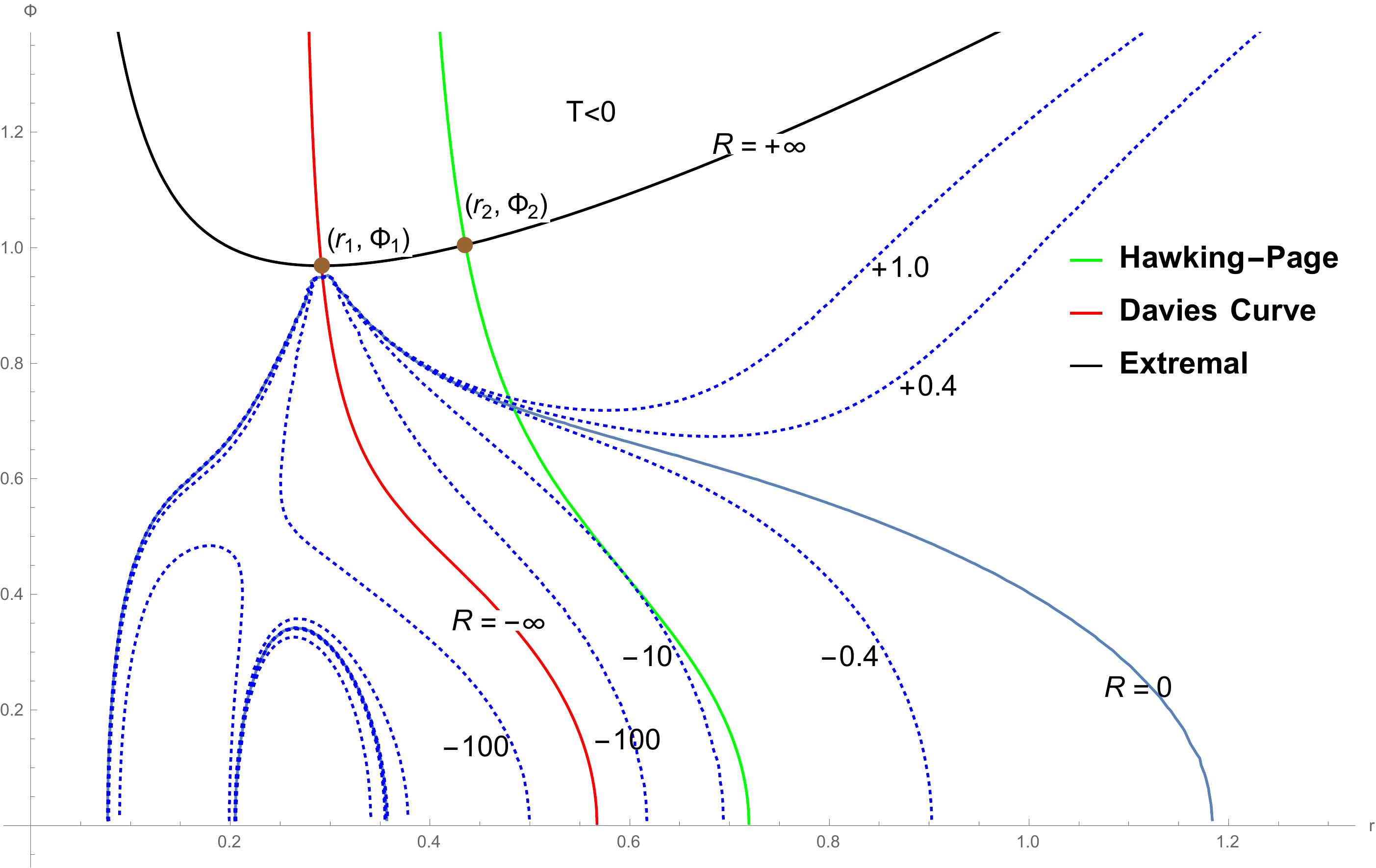}
\caption{Plot in the $r_h-\Phi$ plane of the grand canonical ensemble phase structure and the contours of $R$ for the GB-AdS black holes in $d=6$ for $\alpha=0.0060$. The red curve is the spinodal line, the black one is the extremal curve, the blue one is the locus of vanishing $R$ and the green is the Hawking page curve. The dotted blue curves are contours of $R$ with the values indicated.}
\label{Rsign3}
\end{figure}
where $\mathcal{P}_6$ is a polynomial function of   $\Phi, r$ and $\alpha$. Clearly, the scalar curvature diverges along the Davies point. It can be shown that the divergence is always negative infinity. Away from the Davies point on the stable branch, the scalar curvature $R$ turns to small positive values before finally decaying to zero in high temperature regions far away from the instability. This feature of $R$ is exactly the same as that of the RN-AdS black hole \cite{sahay2,sahay3}. Again, just as for RN-AdS black holes, $R$ diverges to positive infinity at the extremal point whenever it exists.

 The positive divergence of $R$ to infinity at extremality follows exactly the same trend as eq. (\ref{extR}). However, the scaling behaviour of $R$ along the Davies point reveals a pertinent difference with its scaling about the isolated critical point in $d=5$. $R$ does not scale as the inverse of the Gibbs free energy at the Davies point, whose scaling was given above in eq. (\ref{gscaledavies}). Rather, one obtains

\begin{align*}
 R_6^{-1} &\sim (\Phi-\Phi_D)\,\,\mbox{along isotherm, and}\\
 R_6^{-1} &\sim (T-T_D)\,\,\mbox{along isopotential.} \label{Rscaledavies}\numberthis
\end{align*}

It is then a numerical coincidence that scaling behaviours of both $R$ and $G_s$ give the right scaling of a response function like $C_\Phi$. This is seen as follow. The scaling of $G_s\sim (T-T_cr)^{3/2}$ on second derivative gives $C_\Phi\sim (T-T_{cr})^{-1/2}$. At the same time, $R$ goes as $\mathcal{N}_6^{\,-2}$ so that the scaling $R\sim(T-T_{cr})$ also implies the same scaling for $C_\Phi$.

In figure (\ref{Rsign3}) we show in the $\Phi-r_H$ plane the global features of geometry superimposed over the phase structure in grand canonical ensemble of $d=6$ (refer to figure (\ref{phasegrand6})). Much of the behaviour outlined in the preceding can be directly inferred from the figure. The region to the left of the the red coded Davies curve (which is also $R=-\infty$) is the locally thermally unstable region and we shall not be analysing it henceforth. Similar to the $d=5$ case in figure (\ref{Rsign}), the Hawking$-$Page curve intersects the locus of zeroes of $R$ at the top right. So, that for $\Phi$ values above the intersection, $R$ is positive at the HP transition and moreover, it cuts higher and higher valued contours of positive $R$ as the HP curve reaches the extremal curve. This behaviour is again unlike the corresponding RN-AdS case, where the HP curve always remains at small negative $R$ values. Again it seems to suggest that the thermal Gauss$-$Bonnet-AdS at lower temperatures has dominant repulsive interactions much like the fermion gas \cite{ruppblack}. At the same time, again much like its $5-d$ counterpart, for $\Phi$ values below the intersection (hence higher temperatures), the HP curve runs almost parallel to small negative values of $R$ thereby informing us, via the $R$-crossing, that the thermal GB-AdS is very much like the ideal gas, with possibly weak attractive forces. 

In figure (\ref{R66}), we plot $R$ vs. $T$ curves for exactly the same values of $\alpha$ and $\Phi$ as in figures (\ref{svstin6}) and (\ref{gvstin6}). 

\begin{figure}
\includegraphics[width=0.5\textwidth, height=7cm]{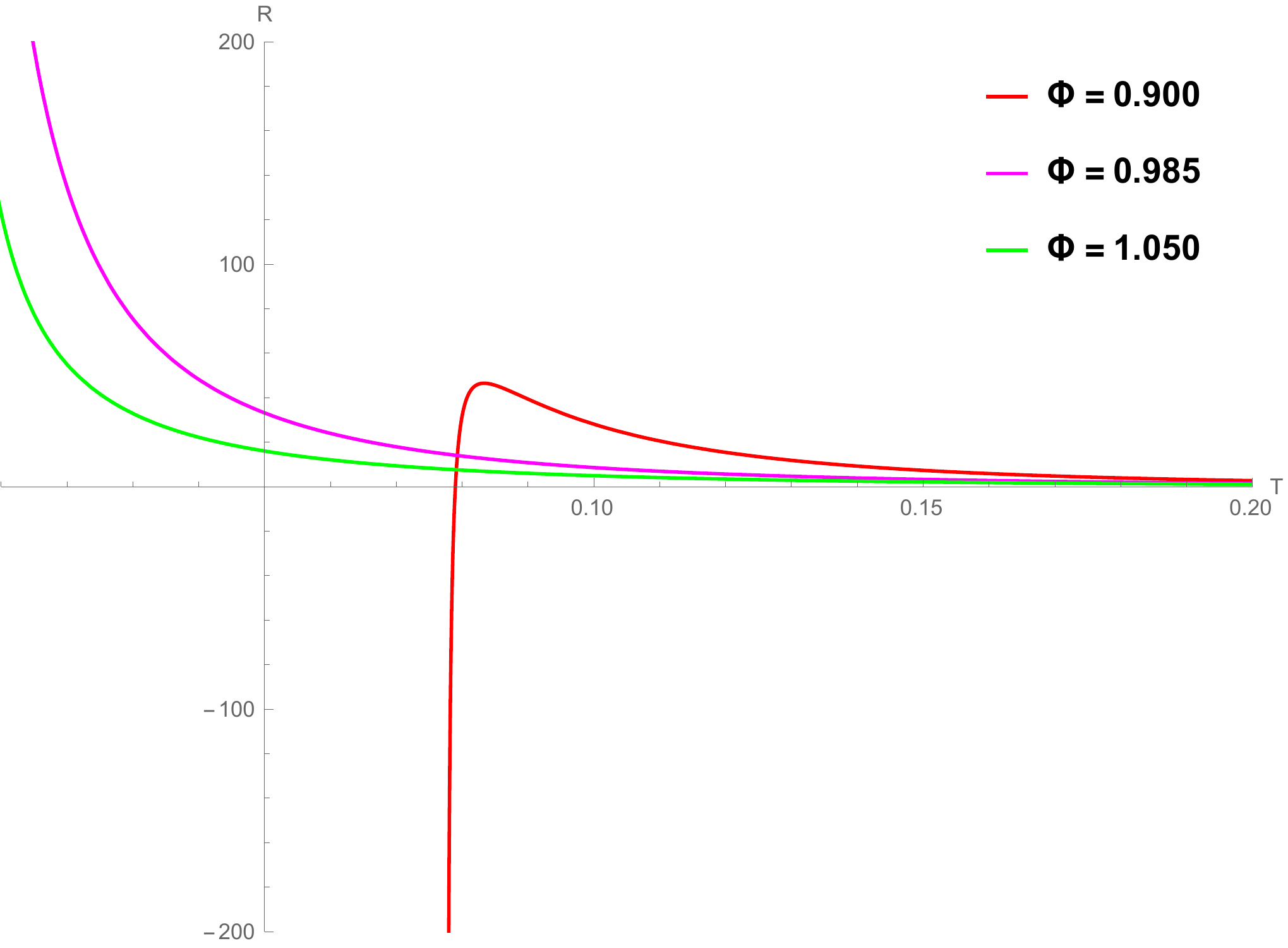}
\caption{Plot of $R$ vs $T$ for the GB-AdS black hole in $d=6$ for $\alpha=0.0060$. The $\Phi$ values are as indicated.}
\label{R66}
\end{figure}

\section{Conclusions}

In this paper, we have undertaken a comprehensive analysis of the phase structure of Gauss-Bonnet-AdS black holes and underlined the role of thermodynamic geometry in explicating the underlying microscopic dynamics associated with coexistence, criticality as well as supercriticality in these black hole systems.

We obtain the state space geometry from the Ruppeiner metric associated with the GB-AdS black holes and find that the thermodynamic manifold encodes the grand canonical ensemble with curvature singularities along the locus of response function infinities. Our analysis near the singular points of the manifold explicitly confirms that near the isolated critical point (for charged GB-AdS in $d=5$), $R$ goes {\it exactly} like the inverse of the singular part of free entropy (modulo a constant). Given that the free entropy or the Massieu function is a ratio of the system size to the correlation volume, the curvature length squared of the associated thermodynamic manifold gets interpreted as the number of uncorrelated microscopic domains of the black hole system. We believe it is a concrete step towards a physical understanding of $R$ for black holes for which the standard interpretation of $R$ as the correlation volume is not immediately applicable. Further, we note that while $R$ diverges to negative infinity, it does not scale as inverse of the free energy near the Davies transition points (found in $d=6$). This naturally leads us to interpret the Davies transitions within the context of the Hawking-Page transition when it is viewed as a first order phase transition between the thermal AdS phase and the black hole phase of the Gauss-Bonnet spacetime. 

We analyze in detail the contours of the scalar curvature, especially its signature, for thermodynamic manifolds in $d=5$ and $d=6$. Several new insights are offered by geometry. A key insight for the $d=5$ case,  where phase coexistence exists, is that the sign of $R$ for the small black hole branch is positive while it is negative for the large black hole branch. This suggests a distinct difference in the underlying microscopic interactions for the two branches. While the small black hole branch has dominant repulsive interactions (and hence is more 'solid like'), the large black hole branch has dominant attractive interactions and is, hence, more 'fluid like'. We equate the magnitude of $R$ in the coexisting regions via the $R-crossing$ method and obtain a coexistence curve without using the mean-field dictated by the Maxwell construction method, which nevertheless closely parallels the one obtained using the latter method. 

Furthermore, geometry helps uncover a distinct supercritical regime in $d=5$. Beyond the critical point, while there is no sharp phase transition, the sign of $R$ and its locus of maxima (in magnitude) together help clearly outline a crossover region between the solid-like-small-black hole and the fluid-like-large-black hole regions. This crossover region, reminiscent of the Widom line in ordinary systems, persists for a while beyond the critical point with decreasing sharpness. We further comment on the positive divergence of $R$ to infinity at extremality for both $d=5$ and $d=6$. While it is plausible that this feature could be due to some mechanical instability at extremality, this could alternatively signal a quantum regime at low temperatures, with the black hole thermodynamics showing similarities with the $2-d$ Fermi gas at low  temperatures \citep{ruppblack}. This then could signal the presence of another class of repulsive statistical interactions different from the solid like interactions for the small black hole branch, namely the Pauli repulsive interactions.
  
  Finally, we discuss an interesting perspective on Hawking-Page transition from information retrieved by the  geometry of the state space in the vicinity of the transition. Viewing the HP transition as a first order transition (see figure (\ref{gvstin6})) between thermal GB-AdS and the locally stable large black hole branch, we use the $R$-crossing method to gain info on the scalar curvature of the thermal AdS near the transition. This reveals a striking difference with the $\alpha=0$ or the RN-AdS case in $d=5$ and $d=6$. For the RN-AdS case at the HP transition, $R$ has consistently order unity negative values suggesting that the thermal GB-AdS phase, or the confined phase via the gauge gravity duality, behaves like an ideal gas with possible weak long range attractive interactions. While there are similar features for non-zero $\alpha$ in $d=5$ and $d=6$ at higher temperatures and lower potential, the low temperature region is a point of significant departure. Thus, for lower temperatures $R$ has large positive values. This suggests a dominant fermionic interaction in the thermal AdS phase around the Hawking-Page transition point.

 \section{Acknowledgement}
 
 AS gratefully acknowledges enlightening discussions with George Ruppeiner during SigmaPhy 2017 and also for several comments on an advance copy of the draft. AS also acknowledges helpful discussions with Ritu Sharma.


\begin{thebibliography}{99}

\bibitem{love}
D. Lovelock, "The Einstein tensor and its generalizations", J. Math. Phys. 12 , 498-501 (1971).

\bibitem{deser}
D. G. Boulware and S. Deser, "String-Generated Gravity Models", Phys. Rev.
Lett. 55, 2656-2660 (1985).

\bibitem{zwei}
B. Zwiebach, "Curvature squared terms and string theories", Phys. Lett. B
156, 315-317 (1985).

\bibitem{malda}
O. Aharony, S. Gubser, J. Maldacena, H. Ooguri, Y. Oz, "Large N Field Theories, String Theory and Gravity", Phys.Rept. 323, 183-386 (2000). [arXiv:hep-th/9905111] 

\bibitem{witt}
E. Witten, "Anti-de Sitter space, thermal phase transition, and confinement in gauge theories", Adv.Theor.Math.Phys. 2, 505-532 (1998). [arXiv:hep-th/9803131]
\bibitem{kovtun}
P. K. Kovtun, D. T. Son, A. O.  Starinets, "Viscosity in strongly interacting quantum field theories from black hole physics", Physical Review Letters. 94, (11), 111601 (2001),  arXiv:hep-th/0405231.

\bibitem{mukund}
M Rangamani, "Gravity and Hydrodynamics, Lectures on the fluid gravity correspondence", Class. Quant. Grav. 26, 224003 (2009), arXiv:0905.4352v3.

\bibitem{subir}
 Subir Sachdev, "From gravity to thermal gauge theories: the AdS/CFT correspondence",  Lectures at the 5th Aegean summer school (2010), arXiv:1002.2947v1.
 
\bibitem{ryu}
T. Nishioka, S. Ryu, T. Takanayagi, "Holographic Entanglement Entropy; A Review", J. Phys. A 42:504008, 2009, arXiv:0905.0932v2.
\bibitem{odin}
S. Nojiri and S. D. Odintsov, "Anti-de Sitter Black Hole Thermodynamics in Higher Derivative Gravity and New Confining-Deconfining Phases in dual CFT", Phys. Lett. B 521, 87-95 (2001) [Erratum {\it ibid}. B542, 301 (2002)].
[arXiv:hep-th/0109122]

\bibitem{sudipt1}
T. K. Dey, S. Mukherji, S. Mukhopadhyay, S. Sarkar, "Phase Transitions in Higher Derivative Gravity", JHEP 0704, 014 (2007). [arXiv:hep-th/0609038]

\bibitem{sudipt2}
T. K. Dey, S. Mukherji, S. Mukhopadhyay, S. Sarkar, "Phase transitions in higher derivative gravity and gauge theory: R-charged black holes", JHEP 0709,
026 (2007). [arXiv:0706.3996 [hep-th]]

\bibitem{thesis}
P. G. Szepietowski,
"Higher Derivative Corrections, Consistent
Truncations, and IIB Supergravity", Doctoral Thesis, The University of Michigan, 2011.

\bibitem{cai}
R. G. Cai, "Gauss-Bonnet black holes in AdS spaces", Phys. Rev. D 65, 084014 (2002). [arXiv:hep-th/0109133]

\bibitem{cho}
 Y. M. Cho and I. P. Neupane, "Anti-de Sitter black holes, thermal phase transition, and holography in higher curvature gravity", Phys. Rev. D 66, 024044 (2002). [arXiv:hep-th/0202140]
 
\bibitem{holoheatengine}
Clifford V. Johnson, "Holographic Heat Engines", Class. Quant. Grav. 31 (2014) 205002. [arXiv:1404.5982 [hep-th]]

\bibitem{chandra}
C. Bhamidipati, P. K. Yerra, "A Note on Gauss-Bonnet Black Holes at Criticality", arXiv:1706.09344v1 [hep-th]

\bibitem{manman}
A M Frassino, D Kubiznak, R B Mann, F Simovc, "Multiple reentrant phase transitions and triple points in Lovelock thermodynamics", JHEP09(2014)080, [arXiv: 1406.7015 [hep-th]]


\bibitem{caicai}
R. G. Cai, S. P. Kim, B. Wang, "Ricci flat black holes and Hawking-Page phase transition in Gauss-Bonnet gravity and dilaton gravity", Phys. Rev. D
76, 024011 (2007). [arXiv:0705.2469]

\bibitem{dumitru}
D. Astefanesei, N. Banerjee, S. Dutta, "(Un)attractor black holes in higher derivative AdS gravity", JHEP 0811:070 (2008).  [arXiv:0806.1334] 

\bibitem{zou}
D. C. Zou, Y. Liu, B. Wang, "Critical behavior of charged Gauss-Bonnet-AdS black holes in the grand canonical ensemble", Phys. Rev. D 90, 044063 (2014). [arXiv:1404.5194v2 [hep-th]]

\bibitem{caicaicai}
R. G. Cai, L. M. Cao, L. Li, R. Q. Yang, "P-V criticality in the extended phase space of Gauss-Bonnet black holes in AdS space", JHEP 1309, 005 (2013). [arXiv:1306.6233 [gr-qc]]

\bibitem{greek}
D. Anninos, G. Pastras, "Thermodynamics of the Maxwell-Gauss-Bonnet anti-de Sitter Black Hole with Higher Derivative Gauge Corrections", JHEP 0907:030 (2009). [arXiv:0807.3478v2]
\bibitem{konop}

  R.~A.~Konoplya and A.~Zhidenko,
  ``Quasinormal modes of Gauss-Bonnet-AdS black holes: towards
holographic description of finite coupling,''
  arXiv:1705.07732 [hep-th].
\bibitem{jan}
J Aman, I Bengtsson, N Pidokrajt, "Geometry of black hole thermodynamics", Gen.Rel.Grav.35:1733, (2003), arXiv:gr-qc/0304015v1
\bibitem{ruppm1}
G. Ruppeiner, "Stability and fluctuations in black hole thermodynamics", Physical Review D 75 (2), 024037, (2007)
\bibitem{rupp0}
G. Ruppeiner, "Thermodynamics: A Riemannian geometric model", Phys. Rev. A 20, 1608 (1979)
\bibitem{rupp}
G. Ruppeiner, "Riemannian geometry in thermodynamic fluctuation theory", Rev. Mod. Phys. {\bf 67} 605 (1995) [Erratum {\it ibid}. {\bf 68} 313 (1996)].

\bibitem{kall}
G. Gibbons, R. Kallosh, B. Kol, "Moduli, Scalar Charges, and the First Law of Black Hole Thermodynamics", Phys. Rev. Lett. 77, 4992 (1996). [arXiv:hep-th/9607108v2]

\bibitem{rupp2}
G. Ruppeiner, A. Sahay, T. Sarkar, G. Sengupta, "Thermodynamic Geometry, Phase Transitions, and the Widom Line", Phys.Rev. E86, 052103 (2012). [arXiv:1106.2270v2 [cond-mat.stat-mech]]
\bibitem{rupp4}
H. May, P. Mausbach, ``Riemannian geometry study of vapor-liquid phase equilibria and supercritical behavior of the Lennard-Jones fluid'', Phys. Rev. E 85, 031201 (2012),  Phys. Rev. E 86, 059905 (2012).

\bibitem{rupp3}
G. Ruppeiner, M. Mausbach, H. O. May, "Thermodynamic R Diagrams Reveal Solid Like Fluid States", Phys. Lett. A
Volume 379, Issue 7, Pages 646–649, 20 March 2015. [arXiv:1411.2872v1 [cond-mat.stat-mech]]

\bibitem{sahay}
A. Sahay, "Restricted thermodynamic fluctuations and the Ruppeiner geometry of black holes", Phys. Rev. D 95, 064002 (2017). [arXiv:1604.04181v3 [hep-th]]

\bibitem{gbc1}
S. Wei, Y. Liu, "Critical phenomena and thermodynamic geometry of charged Gauss-Bonnet AdS black holes", Phys. Rev. D 87, 044014 (2013). [arXiv:1209.1707v2]

\bibitem{dolanT}
B. P. Dolan, "Intrinsic curvature of thermodynamic potentials for black holes with critical points", Phys. Rev. D 92, 044013 (2015). [arXiv:1504.02951v1 [gr-qc]]

\bibitem{cliff}
A. Chamblin, R. Emparan, C. V. Johnson, R. C. Myers, "Holography, thermodynamics, and fluctuations of charged AdS black holes",  Phys. Rev. D 60, 104026 (1999). [arXiv:hep-th/9904197v1]

\bibitem{sahay2}
A. Sahay, T. Sarkar, G. Sengupta, "On the thermodynamic geometry and critical phenomena of AdS black holes", JHEP 1007:082 (2010). [arXiv:1004.1625v2 [hep-th]]
 
\bibitem{sahay3}
A. Sahay, T. Sarkar, G. Sengupta, "On the Thermodynamic Geometry and Critical Phenomena of AdS Black Holes", JHEP 1007:082 (2010). [arXiv:1004.1625v2 [hep-th]]
\bibitem{gubser}
S. S Gubser, "Thermodynamics of spinning D3-branes", Nucl.Phys. B551 (1999) 667-684, arXiv:hep-th/9810225v2
\bibitem{stefano}
S. Aretakis, "Horizon Instability of Extremal Black Holes ", Adv.Theor.Math.Phys. 19 (2015) 507-530, arXiv:1206.6598 [gr-qc]
\bibitem{rupp6}
G. Ruppeiner, "Thermodynamic curvature and black holes", "Breaking of Supersymmetry and Ultraviolet Divergences in Extended Supergravity," Springer Proceedings in Physics Volume 153, 2014, pp 179-203, arXiv:1309.0901v2

\bibitem{mann}
D. Kubiznak, R. Mann, "P-V criticality of charged AdS black holes", JHEP 1207:033 (2012). [arXiv:1205.0559v2]

\bibitem{stan}
L. Xu, P. Kumar, S. V. Buldyrev, S. H. Chen, P. H. Poole, F. Sciortino, H. E. Stanley, "Relation between the Widom line and the dynamic crossover in systems with a liquid–liquid phase transition", PNAS 2005 102: 16558-16562

\bibitem{scop}
G. G. Simeoni. T. Bryk, F. A. Gorelli, M. Krysch, G. Guorro, M. Santoro, T. Scopigno, "The Widom line as the crossover between liquid-like and gas-like behaviour in supercritical fluids", Nature Physics 6, 503–507 (2010).
\bibitem{page}
S. W. Hawking, D. N. Page, "Thermodynamics of Black Holes in Anti-de Sitter Space", Comm. Math. Phys. 87(577-588), 1983. 
\bibitem{majhi}
S. Samanta, S. Mandal, B. R. Majhi, Phys. Rev. D 94, No.6, 064069 (2016). [arXiv:1608.04176 [gr-qc]]
\bibitem{rupp5}
G. Ruppeiner, N. Dyjack, A. McAloon, and J. Stoops, "Solid-like features in dense vapors near the fluid critical point", The Journal of Chemical Physics 146, 224501 (2017)
\bibitem{stauff}


D. Stauffer, M. Ferer, and Michael Wortis, ``Universality of Second-Order Phase Transitions: The Scale Factor for the Correlation Length'',

Phys. Rev. Lett. 29, 345 (1972)

\bibitem{ruppblack}
G. Ruppeiner, `Thermodynamic curvature and phase transitions in Kerr Newman black holes', Phys. Rev. D 78, 024016 (2008)
\bibitem{wein}
S. Weinberg, `Gravitation and Cosmology', ISBN: 978-0-471-92567-5.

\end{thebibliography}
\end{document}